\documentclass[journal]{IEEEtran}

\usepackage{multirow}
\usepackage{diagbox}
\usepackage[hyphens]{url}
\usepackage{hyperref} 

\usepackage{mathtools, algorithm}
\usepackage{amssymb}
\usepackage[caption=false]{subfig}

\usepackage{bbm}

\usepackage{algorithm,algorithmicx,algpseudocode}
\usepackage{amsthm,amsmath}
\usepackage{xcolor}
\usepackage{tabularx,booktabs}
\usepackage{tablefootnote}
\usepackage{listings}

\usepackage[caption=false]{subfig}

\theoremstyle{definition}

\colorlet{punct}{red!60!black}
\definecolor{background}{HTML}{EEEEEE}
\definecolor{delim}{RGB}{20,105,176}
\colorlet{numb}{magenta!80!black}

\usepackage[bottom]{footmisc}
\usepackage{float}
\usepackage{graphicx}
\usepackage{colortbl}
\usepackage{listings}
\usepackage{diffpath}
\usepackage{dblfloatfix}
\usepackage[capitalise,nameinlink]{cleveref}
\usepackage{numprint}
\usepackage{tcolorbox}
\usepackage{etoolbox}

\usepackage[frozencache,cachedir=.]{minted}

\colorlet{punct}{red!60}
\definecolor{background}{HTML}{EEEEEE}
\definecolor{delim}{RGB}{20,105,176}
\colorlet{numb}{magenta!80!black}

\usepackage[T1]{fontenc}
\lstdefinelanguage{json}{
    basicstyle=\normalfont\ttfamily,
    numbers=left,
    numberstyle=\scriptsize,
    stepnumber=1,
    numbersep=8pt,
    showstringspaces=true,
    breaklines=true,
    frame=lines,
    backgroundcolor=\color{background},
    literate=
      {:}{{{\color{punct}{:}}}}{1}
      {,}{{{\color{punct}{,}}}}{1}
      {\{}{{{\color{delim}{\{}}}}{1}
      {\}}{{{\color{delim}{\}}}}}{1}
      {[}{{{\color{delim}{[}}}}{1}
      {]}{{{\color{delim}{]}}}}{1},
}

\usepackage{silence}
\IEEEoverridecommandlockouts

\begin{document}
\title{AUDITEM: Toward an Automated and Efficient Data Integrity Verification Model Using Blockchain}

\author{\IEEEauthorblockN{Zeshun Shi\IEEEauthorrefmark{1},
Jeroen Bergers\IEEEauthorrefmark{1},
Ken Korsmit\IEEEauthorrefmark{2}, and
Zhiming Zhao\IEEEauthorrefmark{1}}

\IEEEauthorblockA{\IEEEauthorrefmark{1}Informatics Institute, University of Amsterdam, Amsterdam, 1098 XH, the Netherlands 
\\\IEEEauthorrefmark{2}Spatial Eye B.V., Culemborg, 4105 JH, the Netherlands 
\\Email: z.shi2@uva.nl, jeroen.bergers@hetnet.nl, ken.korsmit@spatial-eye.com, z.zhao@uva.nl}
}

\markboth{Journal of \LaTeX\ Class Files,~Vol.~14, No.~8, August~2021}%
{Shell \MakeLowercase{\textit{et al.}}: A Sample Article Using IEEEtran.cls for IEEE Journals}

\maketitle
\begin{abstract}
Data tampering is often considered a severe problem in industrial applications as it can lead to inaccurate financial reports or even a corporate security crisis. A correct representation of data is essential for companies' core business processes and is demanded by investors and customers. Traditional data audits are performed through third-party auditing services; however, these services are expensive and can be untrustworthy in some cases. Blockchain and smart contracts provide a decentralized mechanism to achieve secure and trustworthy data integrity verification; however, existing solutions present challenges in terms of scalability, privacy protection, and compliance with data regulations. In this paper, we propose the AUtomated and Decentralized InTegrity vErification Model (AUDITEM) to assist business stakeholders in verifying data integrity in a trustworthy and automated manner. To address the challenges in existing integrity verification processes, our model uses carefully designed smart contracts and a distributed file system to store integrity verification attributes and uses blockchain to enhance the authenticity of data certificates. A sub-module called Data Integrity Verification Tool (DIVT) is also developed to support easy-to-use interfaces and customizable verification operations. This paper presents a detailed implementation and designs experiments to verify the proposed model. The experimental and analytical results demonstrate that our model is feasible and efficient to meet various business requirements for data integrity verification.
\end{abstract}

\IEEEpeerreviewmaketitle

\begin{IEEEkeywords}
Data management, Integrity verification, Decentralization, Blockchain
\end{IEEEkeywords}

\section{Introduction}
\label{introduction}
Information and data are the most valuable assets of industrial companies and need to be handled properly to maximize their business potential \cite{mork2015contemporary}. With the rapid development of Internet technology, more and more data is being produced every day. Accordingly, various data storage and management tools have emerged to support different business operations, including databases, data warehouses, and data lakes. These data management solutions have different architectures and application scenarios. For example, traditional databases are suitable for basic, day-to-day transaction processing. In contrast, data warehouses help decision-makers with real-time decision support by online analytical processing through a unique data storage architecture. Data lakes, on the other hand, can store data in any format and analyze it, often through machine learning or data mining techniques to achieve greater value from the data \cite{datarepository}. In business operation practices, the joint utilization of these tools can provide enterprises with more options and deliver more benefits. However, despite the advantages of using these data management tools, data stored in the source repository can be edited by internal or external people in the system. This leads to the possibility of people tampering with data to cause huge business losses.

Data integrity ensures the correctness and consistency of data throughout its life cycle. It, therefore, plays a vital role in the design, implementation, and utilization of any data management system \cite{boritz2005practitioners}. The current solution for data integrity verification is through third-party auditing service providers such as Spectra \cite{Spectra}. However, these centralized services are not immune to malicious auditors and therefore suffer from a single point of failure (SPOF) \cite{ninglu}. Besides, the expensive commission fees also discourage companies from using these services due to increased operating costs. According to a report from Audit Analytics, a typical company with revenues of \texteuro25 billion would expect audit fees of around \texteuro12.3 million \cite{auditanalytics}. Recently, blockchain and smart contracts have brought promising hints to address the challenges in the data integrity verification process. In general, a blockchain is a decentralized ledger technology with a ``chained blocks'' data structure. Blocks are linked to each other using a cryptographic algorithm, and every block header includes the root of a Merkle tree, a timestamp, and a hash of the previous block \cite{vademecum}. When the data in any block is changed, the hash value is changed accordingly, resulting in rejection by the network. Consequently, once data is committed to the blockchain, it is immutable, transparent, and decentralized. Combining these properties with a data verification model can achieve the following features:

\begin{itemize}
    \item Immutability of the verification certificates. Any data stored in the blockchain is immutable. Therefore, blockchain can create a tamper-proof environment and enhance the trustworthiness of the data verification result \cite{kalis}.
    \item Decentralization of the verification process. There is no need to use third-party auditing services, which avoids cheating auditors and saves a lot of commission fees and operation costs.
    \item Automation of the verification process. Smart contracts can be used as self-executing and automated validation control systems, allowing different stakeholders to easily verify data integrity without tedious manual operations.
\end{itemize}

Correct implementation of blockchain and data management tools can reliably verify data integrity. This can replace the current expensive and untrustworthy third-party audit process and save time and costs \cite{brandon2016blockchain,zhou2018trustworthy}. The potential of using blockchain as data management and verification systems has been investigated in recent studies \cite{9311629,9548672,9698094,8936349}
This paper builds on top of the previous research and leads to creating the AUtomated and Decentralized InTegrity vErification Model (AUDITEM). Unlike existing models, AUDITEM is designed to handle large-scale data stored in classical data repositories while providing integrity verification capabilities. AUDITEM prevents malicious behavior not only from third parties but also from users within the enterprise organization.

\subsection{Motivating Scenario}
One of the possible use cases for AUDITEM is the utility and telecommunication company. This kind of company usually has numerous fixed assets in the form of utility lines and pipes that must be properly stored and managed \cite{metje2020improved}. A widely used solution is the spatial data warehouse, as it provides various functions to track data changes and obtain daily asset information.
For example, one of these functions is the financial reporting of assets, which requires that the integrity of the data is not compromised. Unfortunately, this is not always the case; there are many examples of integrity issues in reporting data assets. The two most famous incidents are the WorldCom and Enron scandals, both leading to millions of debt because of poor financial reporting \cite{brandon2016blockchain}. 
These events led to the introduction of the Sarbanes-Oxley Act (SOX), a U.S. law to protect investors. It proposes that public companies must have a strong external audit of their bookkeeping.
Currently, utility companies usually store sensitive asset data on their private servers. This does not guarantee that the data is not tampered with, requiring comprehensive auditing. However, the auditing task is expensive, and the auditor can be biased. 
The proposed solution uses blockchain as a trustworthy auditing device. A combination of the data warehouse with the blockchain could lead to the required prevention of data fraud expected by the government and investors \cite{brandon2016blockchain}. 

\subsection{Problem Statement}
\label{rqs}
The main objective of AUDITEM is to prove that the data in the data repository is verifiable and authentic. This is necessary for companies and researchers where data assets are crucial for their core business. One of the current solutions is to use a third-party audit, but this is expensive and cannot be fully trusted. Blockchain can help as it is decentralized and tamper-resistant. However, it also faces challenges such as traditional blockchain technology cannot provide sufficient transaction throughput. Besides, the system must prevent data leakage to outside parties as the data can contain private company information. Finally, we must find a solution when personal information is stored in the blockchain. This is because personal data regulations, e.g., the right to be forgotten in General Data Protection Regulation (GDPR), conflict with the blockchain where data is immutable and cannot be deleted. 
These considerations promote the following research questions:
\begin{itemize}
    \item How to design and implement an automated and efficient data integrity verification model using blockchain?
\end{itemize}

To answer this main research question, we further define the following sub-questions: 
\begin{enumerate}
    \item What blockchain technologies are suitable for data verification, and how scalable are these technologies when more data need to be validated?
    \item How to prevent blockchain from sharing private company data with unauthorized parties while still ensuring data integrity?
    \item How to comply with national personal data regulations when personal data is stored in blockchain for data verification?
\end{enumerate}

\subsection{Contributions}
The main contribution of this paper is a novel decentralized model called AUDITEM to verify data integrity in a trustworthy and efficient manner.
We further introduce a second component called the Data Integrity Verification Tool (DIVT). This tool is an addition to the traditional warehouse by automating the entire validation process. In brief, the main contributions of this paper can be summarized as follows: 
\begin{itemize}
    \item A blockchain-based decentralized model for data integrity verification, where users can customize the degree of traceability based on the importance of the data. 
    \item The DIVT protocol, which is a flexible addition on top of the traditional data management system that combines the model, functions, and algorithms into a single tool with user interfaces.
    \item A methodology to verify data integrity via blockchain when privacy regulations (e.g., GDPR) require the data to be deleted. 
    \item A prototype system based on Hyperledger Fabric is fully developed and tested. The experimental results demonstrate that AUDITEM is feasible and effective in verifying data integrity.
\end{itemize}

\begin{table*}[!t]
\centering
\caption{Comparison of existing data verification models and their limitations.}
\label{CompariosTechniques}
\resizebox{\textwidth}{!}{%
\begin{tabular}{@{}lll@{}}
\toprule
\textbf{Ref.} &
  \textbf{Topic} &
  \textbf{Limitations} \\ \midrule
Chen et al. \cite{chen2020dynamic} &
  Dynamic data auditing &
  Still rely on trusted third parties to confirm data integrity. \\
Gan et al. \cite{gan2018efficient} &
  Outsourced big data auditing &
  Still rely on trusted third parties to confirm data integrity. \\
Lu et al. \cite{ninglu} &
  Outsourced data auditing &
  Still rely on trusted third parties to confirm data integrity. \\
Hao et al. \cite{hao2018decentralized} & Data integrity verification & PoW-based permissionless blockchain leads to high costs and limited scalability. \\
Wang et al. \cite{wang2019blockchain} &
  Provable data possession &
  Only works with simple data types. High computation latency for large data files. \\
Li et al. \cite{li2018blockchain} &
  IoT data storage and protection &
  Require radical changes to the existing data management system. \\
Zyskind et al. \cite{zyskind2015decentralizing} &
  Personal data management &
  Only designed for personal profiles. Lack of implementation of the model. \\
Sun et al. \cite{sun2020blockchain} &
  Medical records storage and retrieval &
  Only designed for health care data records. \\
Kalis and Belloum \cite{kalis} &
  Data integrity verification &
  Only high-level tamper-proof traceability. PoW causes high costs and low scalability. \\ \bottomrule

\end{tabular}%
}
\end{table*}

The remainder of this paper is organized as follows: 
In \cref{relatedwork}, the research background is investigated, including a detailed summary of the related work and research gaps.
In \cref{Overview}, the model overview is presented. The section starts with an analysis of the requirements and actors. Then, the system components, process flows, and privacy regulations are described. Next, \cref{implementation} introduces the implementation details of the AUDITEM prototype. In \cref{evaluation}, our model is evaluated with detailed experiments to investigate whether it meets the requirements.
\cref{discussion} discusses the advantages as well as limitations. Finally, in \cref{conclusion}, the paper is concluded, and the future work is presented. 

\section{Related Work}
\label{relatedwork}
This section first reviews the state-of-the-art data integrity verification models related to AUDITEM. Then, existing research gaps are summarized, and the position of this study is discussed.

\subsection{Data Integrity Verification Models}
Existing related research is in multiple directions. Traditional data integrity verification studies mainly focus on innovating new verification algorithms to support diverse verification needs. In this respect, Chen et al. \cite{chen2020dynamic} and Gan et al. \cite{gan2018efficient} proposed two efficient and secure auditing schemes for dynamic data auditing operations. On the other hand, blockchain has recently been proposed as a trustworthy data auditing device.
Lu et al. \cite{ninglu} tried to improve the traditional audit model by combining third-party auditing services with a blockchain system. Their model constructs a trusted system by randomly assigning a user to multiple untrusted third-party auditors through a selection algorithm.
Hao et al. \cite{hao2018decentralized} proposed a system where the blockchain is leveraged as a data auditing device. The system works by creating two functions: \emph{WriteBlock} and \emph{Checkblock}. In the \emph{WriteBlock} function, the encrypted data is outsourced to an untrusted external data source, and a blockchain with a PoW consensus is used to create and confirm the data signature. In the \emph{CheckBlock} function, the state of the blockchain is confirmed with the data on the external data source using the signature.
This is to prove that the data exist and have not been changed. 
Similarly, Wang et al. \cite{wang2019blockchain} create a blockchain scheme that focuses more on efficient cryptographic methods.
To eliminate data storage problems in traditional centralized servers, Li et al. \cite{li2018blockchain} proposed a distributed IoT data storage scheme using blockchain and cetrificateless cryptography.
Zyskind et al. \cite{zyskind2015decentralizing} created a personal data management tool using an off-chain storage solution with a focus on privacy protection. In their model, the data owner controls the permissions of other parties to view the data.
Sun et al. \cite{sun2020blockchain} introduced a system to store and retrieve medical records via IPFS and blockchain securely. The focus of this work is to encrypt the files stored in IFPS and request the decryption key using specific keywords.
Kalis and Belloum \cite{kalis} presented a more simple algorithm. They decide only to store a hash and a unique identifier on the blockchain. New versions of the data are stored under a new identifier. To verify the integrity, the hash is recalculated and matched with the hash on the blockchain. 
In this model, the data owner is considered untrusted, and the audit trail covers all forms of malicious actors.

It should be noted that each of the discussed models has some limitations when combined with a real-life data repository. 
The models in \cite{gan2018efficient,ninglu,chen2020dynamic} all rely on a trusted third party to confirm the data integrity, not addressing the issue of untrustworthy auditors and certificates. 
The model in \cite{hao2018decentralized} uses a permissionless blockchain with the PoW consensus, which leads to high costs and limited scalability.
Similarly, the model proposed by \cite{wang2019blockchain} may suffer from high computation latency when larger files need to be encrypted and stored \cite{gangadevi2021survey}.
Some of the blockchain-based techniques require a complete change of the current data management tools \cite{li2018blockchain}. 
Besides, they are often designed for small files such as personal profiles \cite{zyskind2015decentralizing} or special data formats such as health care records \cite{sun2020blockchain}, and are not suitable for large tables stored in a data warehouse or data lake. 
The technique presented by \cite{kalis} only stores a periodic hash on a public blockchain. The storage of more hashes results in better traceability of the tampering location but increases the storage burden. Furthermore, an increase in information storage would also lead to an increase in cost, limiting the scalability of this technique. In summary, the topics and limitations of the discussed techniques are shown in \cref{CompariosTechniques}.

\subsection{Existing Research Gaps}
\label{gaps}
The first gap that needs to be addressed is the scalability of the blockchain when building a data verification model. The consensus algorithm is the main factor limiting the scalability of current blockchain-based systems \cite{gangadevi2021survey}. Many presented models work with permissionless blockchains that require a compute-intensive PoW consensus; however, this is not feasible for large-scale data storage and verification. On the other hand, even permissioned blockchains may have the scalability issue, as transaction throughput may not be sufficient in some cases \cite{gdprcompliantpersonaldata}. Besides, there are considerable differences in the discussed literature regarding how data manipulation should be located and recovered. For example, the identification of tampering is at the table level in \cite{kalis}, while in \cite{hao2018decentralized} a small part of a file is checked to verify the authenticity. The amount of data stored in a tamper-proof location determines how many tampered details can be located and recovered. However, keeping more data on the blockchain requires greater scalability and storage space. This results in a constant trade-off between traceability/recoverability and scalability/redundancy when using blockchain for data auditing.

Another gap is how to prevent the blockchain from sharing private data with unauthorized parties while still ensuring the integrity of the data. Considering the transparent nature of the blockchain, most researchers do not recommend storing the actual data on the blockchain. As a result, two main techniques can be adopted for protecting data privacy. The first approach is to extract only the hash of the data file and store it on the blockchain, as illustrated in \cite{kalis}, \cite{zyskind2015decentralizing}. This approach allows for fast and secure verification but does not provide sufficient traceability and recoverability. In fact, comparing file hashes only tells the auditor whether the file has been changed, not where it has been modified. Another approach is to store the encrypted data in an off-chain distributed file system for data comparison. For instance, the authors in \cite{wang2018blockchain,sun2020blockchain} use the attribute-based encryption (ABE) protocol for encryption and use IPFS to store encrypted data. This asymmetric encryption provides good security but may not be suitable for large-scale data integrity verification tasks due to the high computational overhead \cite{varsha2014using}. 

Finally, there is a gap in compliance with national data regulations when personal data is stored in blockchain for data verification. Regardless of whether the discussed models are based on permissionless or permissioned blockchain, it is always difficult to modify and delete data once it is uploaded. This can be a huge problem when it comes to personal data regulations like GDPR, which allow users to delete personal private data \cite{gdprcompliantpersonaldata}. 
A frequently used solution is to store personal data off-chain so that it can be deleted or edited at any time. 
However, this is not a solution for checking the integrity of data repositories that contain personal information (e.g., an address register). For example, if one address needs to be deleted, the file's hash value will change and result in an integrity violation. In addition, malicious actors may also use this feature to compromise the data integrity verification system.

To summarize, the following research gaps are identified in the related studies: 

\begin{enumerate} 
\item Some data verification techniques require radical changes to the core system, and they are not suitable for complex data types, e.g., geometric data.
\item Some data verification models still require one or more trusted third parties to confirm the integrity. However, this can cause a SPOF.
\item Existing blockchain-based data verification models do not meet the required scalability. They have high transaction costs due to the use of PoW or have high computational overhead due to the need to send too many transactions.
\item Existing blockchain-based integrity verification tools do not have the option to handle GDPR because they assume that the user is the data owner or that the data does not need to be changed.
\end{enumerate}

This paper builds on top of our previous conference paper \cite{jeroen}. We have extensively extended the model, protocol, evaluation, and discussion, with more than 60\% of the content newly added. Specifically, In \cref{introduction}, we extended the research background and added the motivating scenario and problem statement subsections. In \cref{relatedwork}, we added a detailed review and comparison of the related studies. In \cref{Overview}, we updated the model to include introductions on schemes/algorithms and privacy regulations. In \cref{implementation}, we updated the implementation details with the chaincode demonstration and the explanation of the encryption algorithm. In \cref{evaluation}, we added a new functional experiment subsection and extended another four benchmarks for blockchain scalability experiments. New experimental results and findings are presented. In \cref{discussion}, we added a detailed discussion about the advantages and limitations of the model, which is missing in the conference manuscript. Finally, the conclusion was updated in \cref{conclusion}.

\section{Model Design}
\label{Overview}
In this section, we introduce the model details of AUDITEM. The section starts with an analysis of the requirements and actors. Then, the system components, process flows, and privacy regulations are introduced in detail.

\subsection{Requirement Analysis}
\label{requirement}

The requirements are derived in compliance with \emph{Spatial Eye}\footnote{\url{https://www.spatial-eye.com/}}, a utility company in the Netherlands that creates data warehouse applications. 
However, the possibility of model usage is not limited to utility and telecommunication companies.
The input of AUDITEM is an existing data management tool such as a database, data warehouse, or data lake. In this section, we use a data warehouse as a demonstration example.
The output of AUDITEM is a certificate that proves the integrity of the data. It is stored on the blockchain and provides a trustworthy endorsement of the source data.
In summary, the following requirements are identified for AUDITEM:

\begin{itemize}
    \item \emph{Decentralization}: The model should be decentralized and not rely on trusted third parties to avoid a SPOF.
    \item \emph{Traceability}: The location of a possible manipulation must be determined at least on the table's column level and can be customized to be more precise if necessary.
    \item \emph{Security}: Users' personal data must be encrypted and secured at all times when stored outside of the data storage products.
    \item \emph{Data Types}: The model should be able to handle various data types for data integrity verification.
    \item \emph{Regulations}: The model should comply with the GDPR's right to be forgotten, i.e., personal data can be deleted after the end of the service.
    \item \emph{Overhead}: The computational overhead should be kept to a few minutes or less when a regular data batch is added to the data repository.
    \item \emph{Scalability}: It should be possible to allow at least a few dozen companies in a consortium to use the system and submit blockchain transactions simultaneously.
\end{itemize}

\begin{figure*}[!t]
\centering
\includegraphics[width=.75\textwidth]{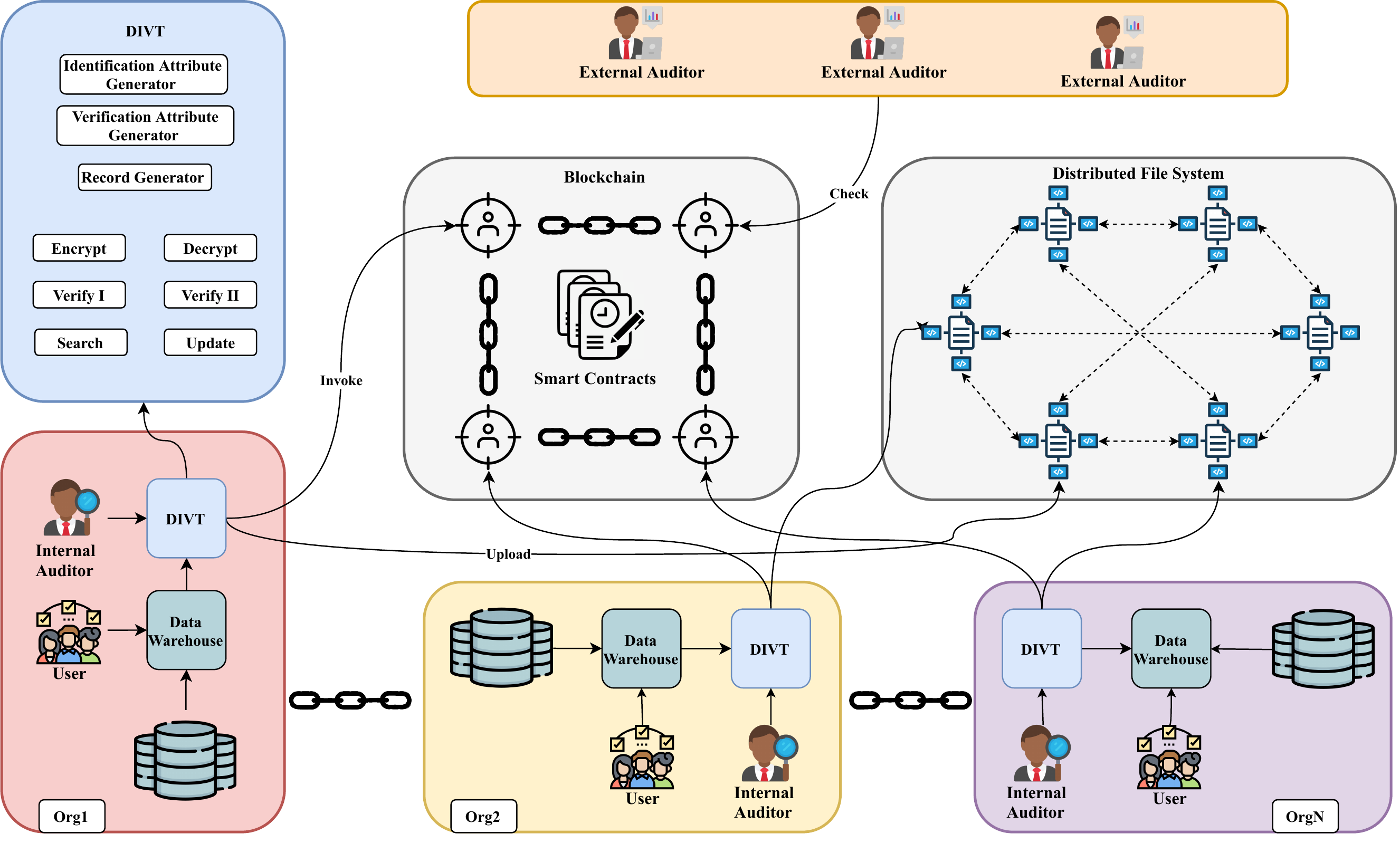}
\caption{Consortium overview of the AUDITEM system.}
\label{architecture}
\end{figure*}

\subsection{Actor Identification}
\label{actor}
The actors and stakeholders that interact with AUDITEM are human actors, external systems, or devices that exchange information with the model. Taking this into account, the following actors are identified:

\begin{enumerate}
\item \emph{Regular User}: The users in the proposed use case are usually analysts who use the geographic information system (GIS). The GIS extracts and uploads data to the data warehouse. Regular users do not want their workflow to be changed; they expect the system to work fully automatic and without performing different operations to verify integrity.
\item \emph{Internal Auditor}: The internal auditor makes the financial reports and expects that the asset data retrieved from the data warehouse is accurate. They do not have much IT knowledge, so the auditing process should be easy to set up. In addition, they expect AUDITEM to be precise, so data mistakes or manipulations can be detected and recovered on time. 
\item \emph{External Auditor}: External auditors can be investors, auditing companies, or government agencies. They expect a company to have perfect bookkeeping. They like to see a certificate to show that everything about the data is correct. It is essential that the system is explainable for external auditors, as they need to trust the certificate. 
\end{enumerate}

\subsection{System Components}
AUDITEM follows a modular design approach and consists of four main components.
A consortium overview of the AUDITEM system is shown in \cref{architecture}. 

\begin{enumerate}
\item \emph{Data Warehouse}: The data warehouse is the data source of the system. It takes care of the retrieval of data stored in the various connected databases. Data is added to the data warehouse in batches on a regular basis.
\item \emph{Data Integrity Verification Tool (DIVT)}: This tool is the communication interface between the system and users. It takes care of basic operations such as encryption, decryption, authentication, and verification and automates the whole process. 
To provide a scalable solution for large-scale data verification,
we design a scheme that combines the Advanced Encryption Standard (AES) and column-level hashes. Besides, two verification methods are designed to support diverse verification requirements.
The schemes and algorithms used in the DIVT protocol are summarized in \cref{fig:protocol3}. The symbolic expressions used in this section are illustrated in \cref{symbol}.
\item \emph{Blockchain}: A consortium permissioned blockchain is leveraged to build trust among different business organizations.
The immutable and decentralized nature of the blockchain makes it suitable for storing and sharing integrity verification hashes and identification attributes.
As more parties join the blockchain network, the system becomes more decentralized with enhanced network protection.
\item \emph{Distributed File System}: A distributed file system is used to take care of the storage of additional verification files because of the limited storage capability of blockchain. The distributed file system should permanently store files and have a precise storage location, preferably based on file content. Options are the InterPlanetary File System (IPFS), Hadoop distributed File System (HDFS), and blockchain-based file systems like StorJ or bigchanDB \cite{huang2020blockchain}. 
\end{enumerate}

\begin{table}[!b]
\centering
\caption{List of symbols used in this section.}
\label{symbol}
\resizebox{\columnwidth}{!}{%
\begin{tabular}{@{}ll@{}}
\toprule
\textbf{Symbol} &
  \textbf{Description} \\ \midrule
$b^y$ & The batch subset $y$ of the data warehouse, added or updated at the same time.  \\
$s_i$ & Identification attribute information where a batch can be identified from. \\
$s_v$ & Verification attribute information that can help with the integrity verification. \\
$h_v$ & The SHA-256 hash of the verification attribute. \\
$r$   & The verification record that contains identification and verification attributes. \\
$p_s$ & The private key used to encrypt and decrypt the verification record $r$. \\ 
$CT$  & The encrypted ciphertext of the verification record $r$. \\
$h_l$ & The location hash of the verification record from the IPFS. \\
\bottomrule
\end{tabular}%
}
\end{table}

\begin{figure*}[!t]
\begin{tcolorbox}[width=\linewidth]

\small
    \parbox{\linewidth}{%
    \begin{center}
    \textbf{Protocol: DIVT}
    \end{center}
    The DIVT protocol is used to assist data users and internal/external auditors to perform automated data integrity verification tasks.
    
    \textbf{Input:} 
    \begin{itemize}
    \item A batch subset of the data warehouse $b^y$. A private key $p_s$ for the encryption/decryption of verification records.
    A blockchain network with configured smart contracts. A distributed file storage system.
    \end{itemize}
    \textbf{Output:} 
    \begin{itemize}
    \item The on-chain certificate indicating whether the data batch holds the integrity.
    \end{itemize}
    \textbf{Functions:} 
    \begin{itemize}
    \item \emph{IDAttGen}. $(b^y) \longrightarrow (s_i$): Identification attribute generator extracts the identification keyword set $s_i$ from the data batch.
    \item \emph{VrfcAttGen}. $(b^y) \longrightarrow (s_v + h_v$): The verification attribute generator extracts the integrity verification attribute $s_v$ from the data batch and generate its hash $h_v$.
    \item \emph{RecGen}. $(s_i + s_v) \longrightarrow (r)$: The record generator is to create a record file $r$ in searchable format based on the verification attribute $s_v$ and identification attribute $s_i$.
    \item \emph{Encrypt}. $(r + p_s) \longrightarrow (CT + h_l)$: The inputs of the encryption function are the verification record $r$ and the private key $p_s$. The created ciphertext $CT$ is stored on the distributed file system, and the location hash $h_l$ is stored on the blockchain.
    \item \emph{Search}. ($s_i ) \longrightarrow (h'_v + h_l$): The identification attribute $s_i$ is used to search the blockchain network via the smart contract to get the new verification hash $h_v'$ and location hash $h_l$.
    \item \emph{Verify I}. $(h_v + h_v')$: The hash $h_v$ created from the data batch is compared with the hash $h_v'$ retrieved from the blockchain. If $h_v = h_v'$, it shows that the data is authentic and has not been tampered with. 
    \item \emph{Decrypt}. $(h_l + CT +p_s) \longrightarrow (r)$: The auditor receives the ciphertext $CT$ from the decentralized storage system using $h_l$, and decrypts it with the private key $p_s$ to receive $r$.
    \item \emph{Verify II}. $(s_v + s_v')$: The verification attribute $s_v$ from the data batch is compared with $s'_v$ received from the distributed file system. If $\forall x$, $s_{v_x} = s'_{v_x}$, then it can be concluded that the data batch is authentic. Otherwise the data is reported as tampered. 
    \item \emph{Update}. $(r) \longrightarrow (r')$: Compare the old verification attribute $s_v$ with the new verification attribute $s'_v$; if changes are allowed, the new verification record $r'$ is created.
    \end{itemize}
    }
\end{tcolorbox}
\caption{The schemes/algorithms used in the DIVT protocol.}
\label{fig:protocol3}
\end{figure*}

\begin{figure}[!b]
\centering
\centering
\includegraphics[width=\linewidth]{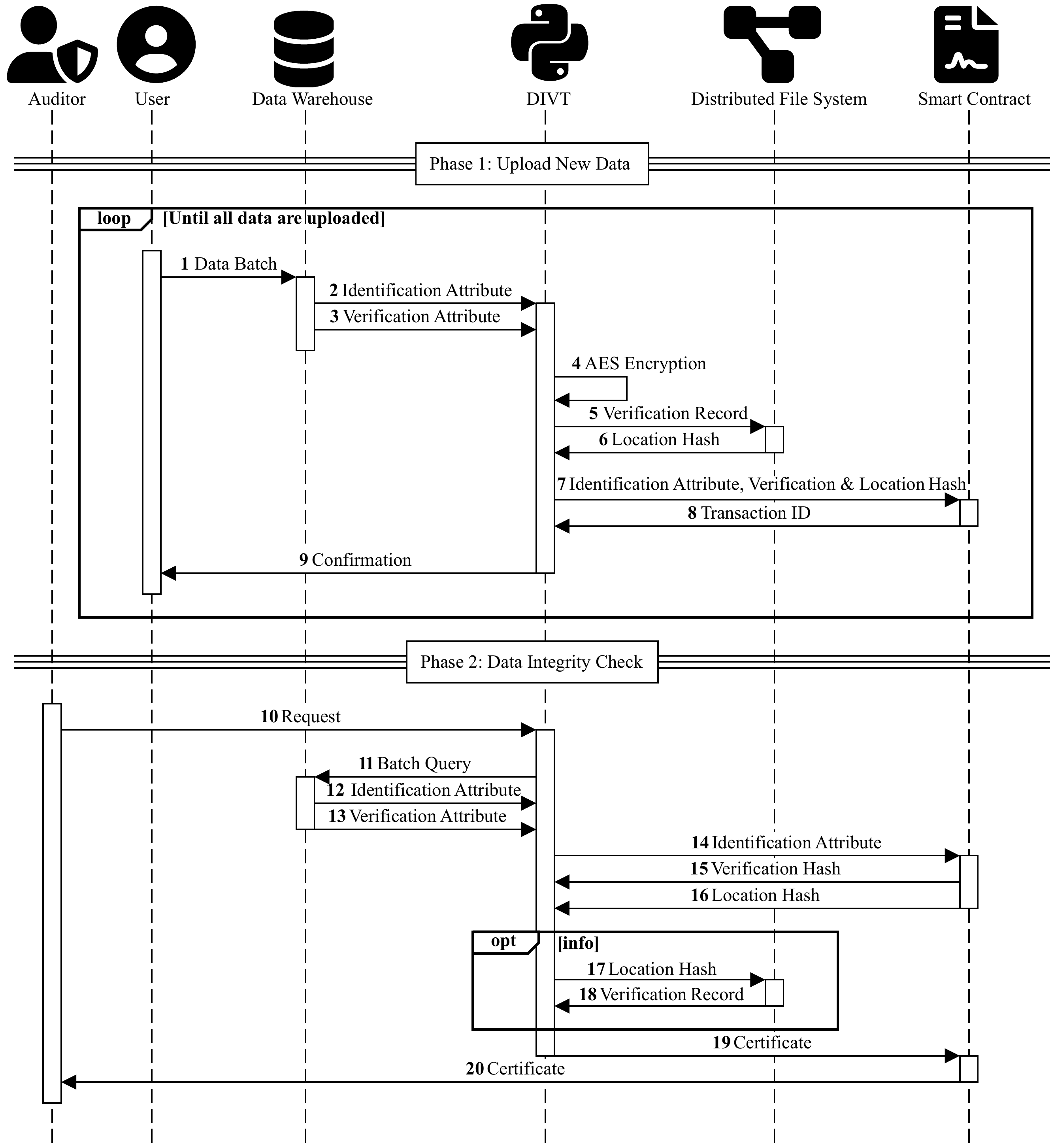}
\label{upload_data}
\caption{Sequence diagram of the AUDITEM process flows.} 
\label{sequence}
\end{figure}

\subsection{Process Flows}
The process flow of AUDITEM can be divided into two general steps: 1) upload new data to the data warehouse and generate attributes, and 2) check data integrity by internal and external auditors.

\subsubsection{Upload New Data}
In the \emph{Spatial Eye} use case, a new network cable is buried in the ground, and information about this cable is sent to the network architect, who later adds it to the GIS along with other changes in the electricity network. Then, the GIS uploads the network changes as batch number 100 to the data warehouse at the end of the day. The DIVT builds on top of the data warehouse extracts identification and verification attributes from this batch. The identification attribute $s_i$ is the keyword set to identify the batch, in this case it can be: $s_i = \{TableId{\,:\,} LowVoltage,\\
\:BatchId{\,:\,}100,\:Timestamp{\,:\,}\operatorname{2017-12-09},\: ... ,\: ... \}$. The verification attribute $s_v$ is to verify the integrity of the warehouse batch and can include the vertical column hashes of the table. In practice, $s_v$ can vary between data tables based on the importance of the data. It is recommended to have at least one SHA-256 hash for each column in $s_v$ to support good traceability. Nevertheless, it should at least consist of a hash of the subset and a description of what other attributes are included. Both $s_i$ and $s_v$ are added together into a verification record $r$, namely $r = s_v + s_i$. $r$ is encrypted using a private key $p_s$, and the returned cyphertext is stored on the distributed file system for later verification. 
A location hash $h_l$ is returned from the distributed file system; this location hash is stored together with verification hash $h_v$ and the identification attribute $s_i$ by broadcasting a transaction to the blockchain smart contract. 
The private key $p_s$ is stored on a private ledger with the identification attribute. An overview of the process can be seen in the first phase of \cref{sequence}.

\subsubsection{Data Integrity Check}
When internal auditors request a data verification task, the DIVT protocol helps them automate this process. An overview of the process flow can be seen in the second phase of \cref{sequence}. First, the auditor selects the batch of data to be validated. For this example, the \emph{LowVoltage} table is used. DIVT then extracts the identification attribute $s_i$ and verification attribute $s_v$ based on the batch number from the data source. Next, the identification attribute $s_i$ is used to search the blockchain network via the smart contract to get $h_v'$ and $h_l$. $h_v'$ retrieved from the blockchain can be compared to $h_v$ received from the batch data with $s_v$. If both are the same value, we can confirm the integrity of this specific batch. However, if the hashes are different, a more advanced verification method is invoked to perform a deep comparison. At this time, DIVT will extract $r$ from the distributed file system and private key $p_s$ from the private blockchain ledger. $s_v'$ obtained from $r$ decryption can be compared with $s_v$ extracted from the \emph{LowVoltage} table to find the differences. A certificate containing the date, auditor, and result is broadcast to the blockchain if the verification of all columns is successful. Otherwise, the modified columns will be reported.
External auditors can confirm a company's data records by checking the certificates on the blockchain when they have the proper permission.

\subsection{Privacy Regulations}
There should be a method to take care of personal data stored in the verification process. An example of this use case could be that in the previously discussed batch 100 of the \emph{LowVoltage} table, some records contain information about the endpoint users. After a year, the service is finished, and the company decides that this data is no longer needed, so it should be deleted. However, if this data is deleted from the warehouse, it will no longer correspond with the blockchain records, resulting in a suggested loss of integrity. A reasonable solution is to create a new verification record and update $h_l$ and $h_v$ on the blockchain. However, this could be a potential vulnerability for the system, i.e., a malicious actor can also delete the data and update the verification records.
To prevent the risks associated with this function, certain countermeasures need to be designed: 1) Only GDPR-sensitive data is allowed to be edited, and these data should not be stored on the blockchain; 2) The changes have to be logged on the blockchain so modifications can be audited and traced; and 3) It has to be easy or automatic to delete the personal data when there is a reason.

To fulfill the above requirements, 
the proposed method lets the smart contract compare the old verification attribute $s_v$ with the new verification attribute $s_v'$. In the record description, $r$ is specified which attributes are essential and which are not. For example, the length of a low voltage cable can never be erased, so $s_{lenght}$ must always be the same as $s_{lenght}'$. However, $s_{EndPoint}'$ is allowed to replace $s_{EndPoint}$ if there is a reason, and the user is allowed to perform this action. As a final precaution, a unique client can be specified within the smart contract. This client can control a GDPR cleanup script that automatically removes GDPR-sensitive data and updates related attributes.

\section{Implementation}
\label{implementation}
This section describes the implementation and technical details of AUDITEM. AUDITEM can be developed using a variety of existing technologies. Our implementation is open-sourced in the Github repository\footnote{\url{https://github.com/ZeshunShi/AUDITEM}}.

\subsection{Design Choices}
The design choices are made in compliance with the utility company \emph{Spatial Eye}. However, there could be more suitable technologies available for other use cases.

\begin{enumerate}
\item \emph{Spatial Data Warehouse}:
The data warehouse product used in this implementation is the \emph{Spatial Warehouse}\footnote{\url{https://www.spatial-eye.com/product-spatial-warehouse.html}}. 
This tool is used to work with numeric and alphabetic data. It can also handle advanced data structures, such as geographic data.
The advantage of this tool is that it can track changes in data over time and facilitate data analysis.

\item \emph{DIVT}:
The DIVT protocol is programmed using the Python Jupyter notebook. Jupyter notebook is an open-source, web-based interactive computational environment\footnote{\url{https://jupyter.org/}}. It is often used by data scientists working in Python or R. 
The decision to use Jupyter notebooks is based on the fact that it provides an easy-to-use interface and interactive data science environment for users.

\item \emph{IPFS}:
For the distributed file system, IPFS is chosen in this implementation\footnote{\url{https://ipfs.io/}}. IFPS meets all our requirements because it is decentralized and allows for faster and secure storage at a lower cost. Besides, it uses a hash over the file's content as a file location. This gives the advantage that data cannot be stored twice as this will result in the same identification hash, saving redundant storage space \cite{sun2020blockchain}.

\item \emph{Hyperledger Fabric}:
Hyperledger Fabric is an open-source permissioned blockchain platform. The reason for choosing a permissioned blockchain is that companies work together in a consortium in the proposed model.
In addition, the efficient consensus mechanism used in the permissioned blockchain can improve scalability and reduce energy waste.
AUDITEM is built with Hyperledger Fabric v2.2, where the recommended consensus protocol is Raft. This consensus mechanism is a Byzantine fault-tolerant mechanism that determines new transactions by a selected leader node \cite{hyperledger}. 
\end{enumerate}

\subsection{Chaincodes Implementation}
The smart contract in Hyperledger Fabric is called chaincode and can be written in programming languages such as Go, Node.js, and java. In order to achieve the model function, the following three chaincodes are designed in AUDITEM.

\usemintedstyle{tango}

\begin{listing}[!t]
\begin{minted}[frame=none,
              breaklines,
              linenos=false,
              xleftmargin=0pt,
              fontsize=\scriptsize,
              bgcolor=black!5,
              tabsize=4]{go}
// Data structure of the verification evidence
type VerificationEvidence struct {
    Organisation string `json:"Organisation"`
    Table_name string `json:"Table_name"`
    Batch_ID string `json:"Batch_ID"`
    Verification_Hash string `json:"Verification_Hash"`
}
// To create the verification evidence in the blockchain
func (s *SmartContract) createEvidence(APIstub shim.ChaincodeStubInterface, args []string) sc.Response {
    var rec = VerificationEvidence{Organisation: args[1], Table_name: args[2], Batch_ID: args[3], Verification_Hash: args[4]}
    RecordAsBytes, _ := json.Marshal(rec)
    APIstub.PutState(args[0], RecordAsBytes)
    compositeKey, err := APIstub.CreateCompositeKey("owner_key", []string{rec.Organisation, args[0]})
    if err != nil {
        return shim.Error(err.Error())
    }
    value := []byte{0x00}
    APIstub.PutState(compositeKey, value)
    return shim.Success(RecordAsBytes)
}
// To query the verification evidence 
func (s *SmartContract) queryEvidence(APIstub shim.ChaincodeStubInterface, args []string) sc.Response {
    RecordAsBytes, _ := APIstub.GetState(args[0])
    return shim.Success(RecordAsBytes)
}
\end{minted}
\caption{Integrity verification chaincode.} 
\label{chaincode1}
\end{listing}

\usemintedstyle{tango}
\begin{listing}[!t]
\begin{minted}[frame=none,
              breaklines,
              linenos=false,
              xleftmargin=0pt,
              fontsize=\scriptsize,
              bgcolor=black!5,
              tabsize=4]{go}
// To create the private key
func(s * SmartContract) createPrivateKey(APIstub shim.ChaincodeStubInterface, args[] string) sc.Response {
  type TransientInput struct {
    SecretKey string `json:"secretKey"`
    Nonce string `json:"nonce"`
    Key string `json:"key"`
  }
  transMap, err := APIstub.GetTransient()
  if err != nil {
    return shim.Error(err.Error())
  }
  privateDataAsBytes, errr := transMap["keys"]
  if !errr {
    return shim.Error("")
  }
  var Input TransientInput
  err = json.Unmarshal(privateDataAsBytes, &Input)
  if len(Input.SecretKey) == 0 {
    return shim.Error("")
  }
  if len(Input.Nonce) == 0 {
    return shim.Error("")
  }
  PrivateDetails := &PrivateDetails {
    SecretKey: Input.SecretKey,
    Nonce: Input.Nonce
  }
  PrivateDetailsAsBytes, err := json.Marshal(PrivateDetails)
  err = APIstub.PutPrivateData("collectionPrivateDetails", Input.Key, PrivateDetailsAsBytes)
  return shim.Success(PrivateDetailsAsBytes)
}
// To query the private key
func(s * SmartContract) queryPrivateKey(APIstub shim.ChaincodeStubInterface, args[] string) sc.Response {
  keyAsBytes, err := APIstub.GetPrivateData(args[0], args[1])
  if err != nil {
    jsonResp := "{\"Error\":\"" + args[1] + err.Error() + "\"}"
    return shim.Error(jsonResp)
  } else if keyAsBytes == nil {
    jsonResp := "{\"Error\":\"" + args[1] + "\"}"
    return shim.Error(jsonResp)
  }
  return shim.Success(keyAsBytes)
}
\end{minted}
\caption{Private storage chaincode.} 
\label{chaincode2}
\end{listing}

\begin{enumerate}
\item \emph{Integrity Verification}: The integrity verification chaincode is the main chaincode to create verification evidence on the blockchain and query them when needed. It includes two main functions: \emph{createEvidence} and \emph{queryEvidence}. The demonstration code is shown in \cref{chaincode1}. First of all, the data structure of verification evidence contains four string fields, namely organisation, table name, Batch ID, and verification hash. In the Hyperledger Fabric blcockchain, write and read operations are performed using the \emph{PutState()} and \emph{GetState()} APIs. \emph{createEvidence} function first reads the verification evidence and encodes the data into a JSON object. Next, the \emph{createCompositeKey()} API is used to create a composite key for the \emph{putState()} function. Finally the record and composite key index are stored separately using the \emph{putState()} API. In the \emph{queryEvidence} function, the record is queried by the \emph{GetState()} API. If the operation is successful, the function returns no error. 

\item \emph{Private Storage}: Hyperledger Fabric offers the ability to create private data collections to store the company's private data, which cannot be viewed by other participants in the blockchain network without permission. In AUDITEM, private data collection is used to store the keys necessary to decrypt the integrity verification files in IPFS. With private data collection, keys can be easily shared with external auditors when needed. The private storage chaincode contains a \emph{createPrivateKey} and a \emph{queryPrivateKey} function, as shown in \cref{chaincode2}. The \emph{createPrivateKey} function first checks if the input parameters are correct, and then the secret key and nonce are saved using the \emph{PutPrivateData()} API. Correspondingly, \emph{queryPrivateKey} function uses the \emph{GetPrivateData()} API to retrieve the data. In Hyperledger Fabric, these two APIs allow smart contracts to store and retrieve private data collections in a secure and privacy-preserving manner.

\item \emph{Integrity Certificates}: The integrity certificate chaincode is responsible for the storage of the certificate that shows the authenticity of the data. It includes a \emph{createCertificate} function, which is invoked when data is identified as authentic, and a \emph{queryCertificate}, which is called by external auditors to query the certificate. The certificate attributes of this chaincode also contain user and time information to facilitate validation.
\end{enumerate}

\subsection{Communications}

\begin{figure}[!t]
\centering
\includegraphics[width=\columnwidth]{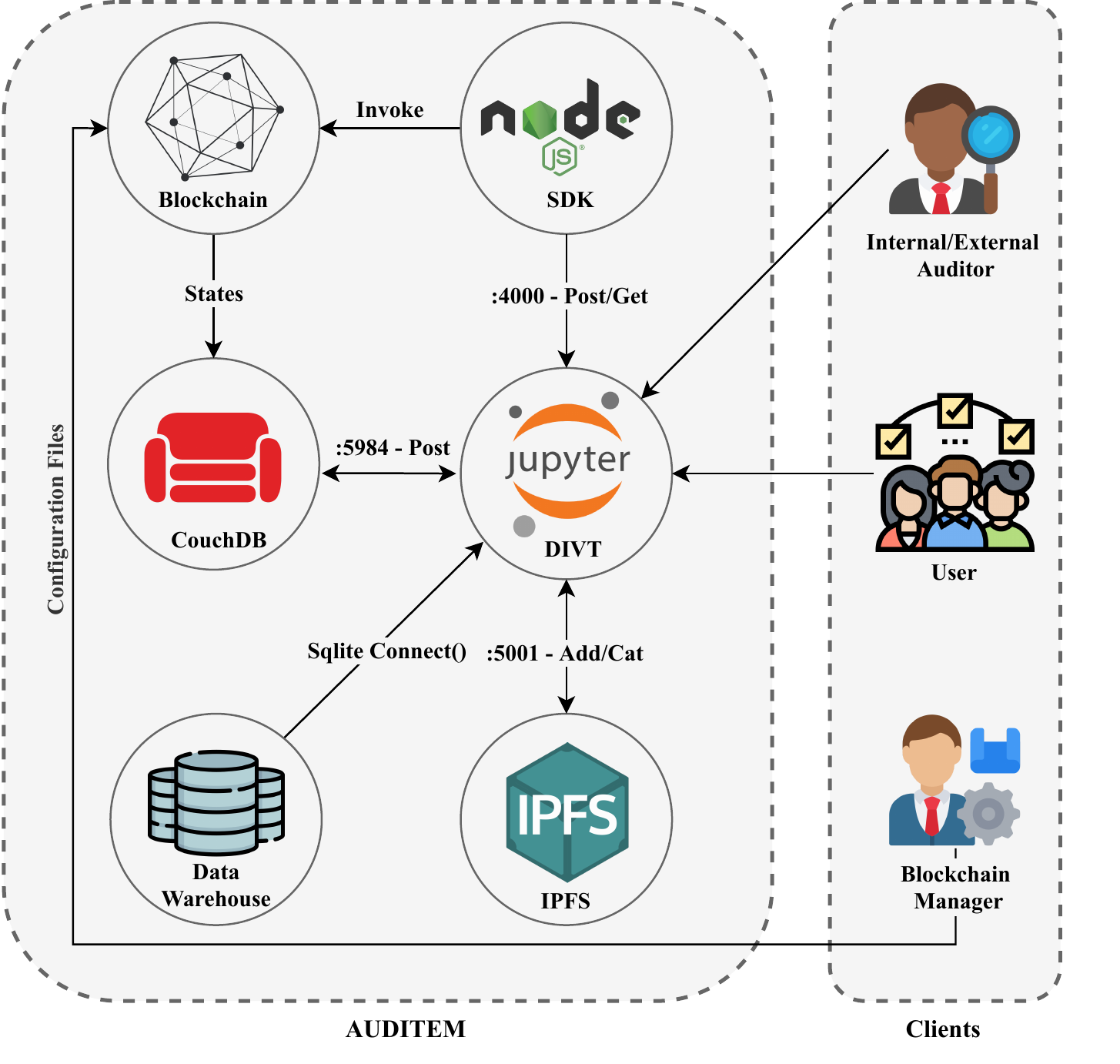}
\caption{Overview of the components and the internal communications.}
\label{internalcom}
\end{figure}

The communication between different components is performed through the REST protocol. This allows using HTTP operations such as GET and POST for component interaction. An overview of the communication within the network can be found in \cref{internalcom}. 
A blockchain manager is needed to configure the blockchain parameters in advance. CouchDB is a state database in the Hyperledger Fabric blockchain that models the ledger data in JSON format. Both CouchDB and IPFS support the possibility of sending HTTP requests via their APIs, which means that operations can be sent using the same logic. The Hyperledger Fabric SDK is used to interact with the blockchain and the DIVT protocol. Especially, it is responsible for checking authorizations, invoking transactions, and querying states. 

\begin{enumerate}
\item \emph{Authorization}:
In Hyperledger Fabric, users are assigned through a certificate authority (CA). The CA distributes certificates to users that can be used to sign and authenticate transactions. For this implementation, we give each company a CA and use the default elliptic curve digital signature algorithm (ECDSA) \cite{fabric_ecdsa} to generate user certificates. 
Besides, the JSON web token (JWT) is used to sign REST API requests and invoke transactions. For example, a new user can be created using a \emph{Post} request on the registered URL. The system returns a JWT token that can be used as a bearer token for the organization's transactions. 
\item \emph{Invoke}:
In order to invoke a transaction to the blockchain, a \emph{Post} request needs to be made to the SDK. The \emph{Post} request must mention the chaincode and name of the chaincode function to invoke a transaction. The SDK uses the JWT token to determine if the user has the proper authorization. The payload of the transaction is a JSON string; for private data, this is transient. 
\item \emph{Query}:
The query operation happens via the same logic as the invoke operation but is used when a \emph{Get} request is received. Their difference is that the invoke operation is used for transactions that change the blockchain state. Instead, the query operation only queries the state database and does not create a new transaction or block.
\end{enumerate}

\subsection{DIVT Protocol}
The DIVT protocol is the automated interface for users to interact with the data warehouse, blockchain, and IPFS. In addition, it is responsible for creating and comparing verification attributes and sending requests to the blockchain to generate certificates.
We have identified three types of AUDITEM users in \cref{actor}. Each type of user corresponds to a different system function and requires a unique user interface. Therefore, three notebook interfaces are created in the DIVT protocol: one for uploading data to the warehouse by regular users, one for checking integrity by internal auditors, and one for checking integrity certificates by external auditors. In the following text, we introduce the key techniques used in the DIVT protocol.

\begin{enumerate}
\item \emph{Verification Options}:
When the database is loaded, individual verification files can be created based on the number of batches and tables in the database. Verification attributes are JSON files containing different fields that can be customized according to the importance of the data. Storing more verification attributes increases the traceability and recoverability of the repository. It is important to declare in the JSON file what the properties are and how they are used because the attributes must be recreated for the verification tasks. In addition, for GDPR compliance, it is necessary to know which columns are allowed to be changed. Taking this into consideration, we propose three different verification options in \cref{threeverificationjson}. The three options from top to bottom have sequentially higher traceability, but at the same time, lead to greater computational overhead.
Other configurations can be easily extended to support more complex data types such as images.
\usemintedstyle{tango}

\begin{listing}
\begin{minted}[frame=none,
               linenos=false,
               xleftmargin=0pt,
               fontsize=\scriptsize,
              bgcolor=black!5,
               tabsize=4]{json}
{
  "h_v":"SHA-256 over subset",
  "traceability":"1",
  "cols":"all column names",
  "rows":"number of rows",
  "gdpr":"column names containing GDPR data",
  "gdprHash":"SHA-256 over the subset without GDPR columns",
  "colHash":"vertical calculated SHA-256 for each column"
}

{
  "h_v":"SHA-256 over subset",
  "traceability":"2",
  "cols":"all column names",
  "rows":"number of rows",
  "gdpr":"column names containing GDPR data",
  "gdprHash":"SHA-256 over the subset without GDPR columns",
  "colHash":"vertical calculated SHA-256 for each column",
  "rowHash":"horizontal calculated SHA-256 for each column"
}

{
  "h_v":"SHA-256 over subset",
  "traceability":"3",
  "cols":"all column names",
  "rows":"number of rows",
  "gdpr":"column names containing GDPR data",
  "gdprHash":"SHA-256 over the subset without GDPR columns",
  "data":"all data inside subset"
}
\end{minted}
\caption{JSON data structure of three verification options.} 
\label{threeverificationjson}
\end{listing}
\item \emph{Verification Functions}:
We developed two functions to address different verification tasks. \emph{Verify I} is supposed to be a faster verification method. It uses the verification hash from the blockchain and compares it with the verification hash created from the data batch. If the two hashes are the same, it means that the data has not been changed, and there is no need for deeper investigation. By contrast, \emph{Verify II} is a deeper comparison, as it compares the two integrity files. This takes more time as the file needs to be retrieved from IPFS and compared with the current database state. Therefore, \emph{Verify II} is only used when \emph{Verify I} results that there is something wrong. 

\begin{figure}[H]
\centering
\begin{tikzpicture}[scale=0.25]

  \draw[thick, dashed,] (-1,-4) rectangle (29,7) node[above left] {AES Round Function};
  \draw[thick,->] (-3,2) -- ++(3,0);
  \draw[thick,->] (28,2) -- ++(3,0);
  
  \defineStateSize{4}
  \defineArrowLength{2}
  \defineTextSize{0.4cm}
  \defineTextArrowSize{0.3cm}
  \defineXGap{0}
  \defineYGap{5}
  \defineStateIndex{0}
  \defineLinkRound{1}
  \initdiffpath

 \colorstate{
    ....
    ....
    ....
    ....
  }{}

  \draw[->, thick,shift={(7,-0.5)}] (0.5, 4) 
    -- ++(0, 1.5) 
    -- node[above] {S-box} ++(6, 0) -- ++(0, -1.5);
    
\fill[color=black!15] (6,0) rectangle ++(4,4);
\fill[color=cyan] (7,3) rectangle ++(1,1);

  \colorstate{
    .*..
    ....
    ....
    ....
  }{}

\fill[color=black!15] (12,0) rectangle ++(4,4);
\fill[color=cyan] (13,3) rectangle ++(1,1);
\fill[color=red] (15,0) rectangle ++(1,4);

  \colorstate{
    .*.x
    ...x
    ...x
    ...x
  }{}
  
\fill[color=black!15] (18,0) rectangle ++(4,4);
\fill[color=black!15] (24,0) rectangle ++(4,4);

\fill[color=green] (21,0) rectangle ++(1,4);
\fill[color=red] (18,0) rectangle ++(1,1);
\fill[color=red] (19,1) rectangle ++(1,1);
\fill[color=red] (20,2) rectangle ++(1,1);

  \fill[color=green] (27,0) rectangle ++(1,4);
  \draw[->, thick,shift={(21,0)}] (0.5, 4) 
    -- ++(0, 1) 
    -- node[above] {$\mathbf{M}\times C$} ++(6, 0) 
    -- ++(0, -1);

  \colorstate{
    ...x
    ..x.
    .x..
    x...
  }{}

  \colorstate{
    ....
    ....
    ....
    ....
  }{0}
  
  \path ( 2,-1) node {$s_{0}$};
  \path ( 8,-1) node {$s_{1}$};
  \path ( 14,-1) node {$s_{2}$};
  \path ( 20,-1) node {$s_{3}$};
  \path ( 26,-1) node {$s_{4}$};

  \path ( 5,-2.5) node[draw=black!15,scale=0.7] {AddRoundKey};
  \path (11.5,-2.5) node[draw=cyan,scale=0.7] {SubBytes};
  \path (17,-2.5) node[draw=red,scale=0.7] {ShiftRows};
  \path (24,-2.5) node[draw=green,scale=0.7] {MixColumns};

\end{tikzpicture}
\caption{Overview of the AES encryption used in DIVT.}
\label{aes}
\end{figure}
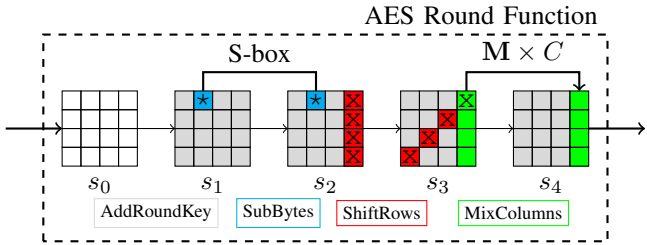

\item \emph{Encryption and Decryption}:
Verification records must be encrypted before uploading to the IPFS. The AES is used in this implementation for encryption and decryption operations in the DIVT protocol. It is a standard chosen by the U.S. government to protect classified information and has been widely used for electronic data protection due to its security and efficient properties \cite{burr2003selecting}. 
The AES encryption algorithm requires multiple runs of the round function. In each round there are generally four operations: \emph{AddRoundKey}, \emph{SubBytes}, \emph{ShiftRows}, and \emph{MixColumns}, as shown in \cref{aes}.
In the \emph{AddRoundKey} step, each byte in the data matrix will perform an XOR operation with the round key derived from the master key.
Next, the main purpose of the \emph{SubBytes} step is to complete the mapping of one byte to another via a substitution box (S-box).
\emph{ShiftRows} is the permutation between bytes within a matrix and is used to provide diffusivity to the algorithm.
\emph{MixColumns} is achieved by multiplying the state matrix with a fixed matrix to reach a diffusion over the columns.
It should be noted that the AES algorithm is a symmetric-key algorithm, meaning that the same key is used for encryption and decryption.
During the encryption, the file needs to be encoded into bytes at first. Then, the file is encrypted using a secret key and a randomly created nonce. Both the secret key and the nonce is stored in the private ledger of the Hyperledger Fabric blockchain. The encrypted data message is then sent to the IPFS for later validation. For decryption, the nonce and secret key can be extracted from the private ledger if the user is from a licensed organization.

\end{enumerate}

\section{Experimental Evaluation}
\label{evaluation}
This section conducts extensive experiments to validate the proposed model. Our tests are divided into functional and non-functional tests. For functional testing, AUDITEM is evaluated to see whether it can accomplish verification tasks. Non-functional tests are further divided into computational overhead and blockchain scalability experiments to test whether AUDITEM is efficient enough for verifying the integrity.

\subsection{Functional Evaluation}
To evaluate the verification capabilities of AUDITEM, two scenarios are designed in which AUDITEM should identify different data manipulation activities. These scenarios use real datasets and have been validated by company experts. 

\begin{figure}[!b]
\centering
\includegraphics[width=0.45\textwidth]{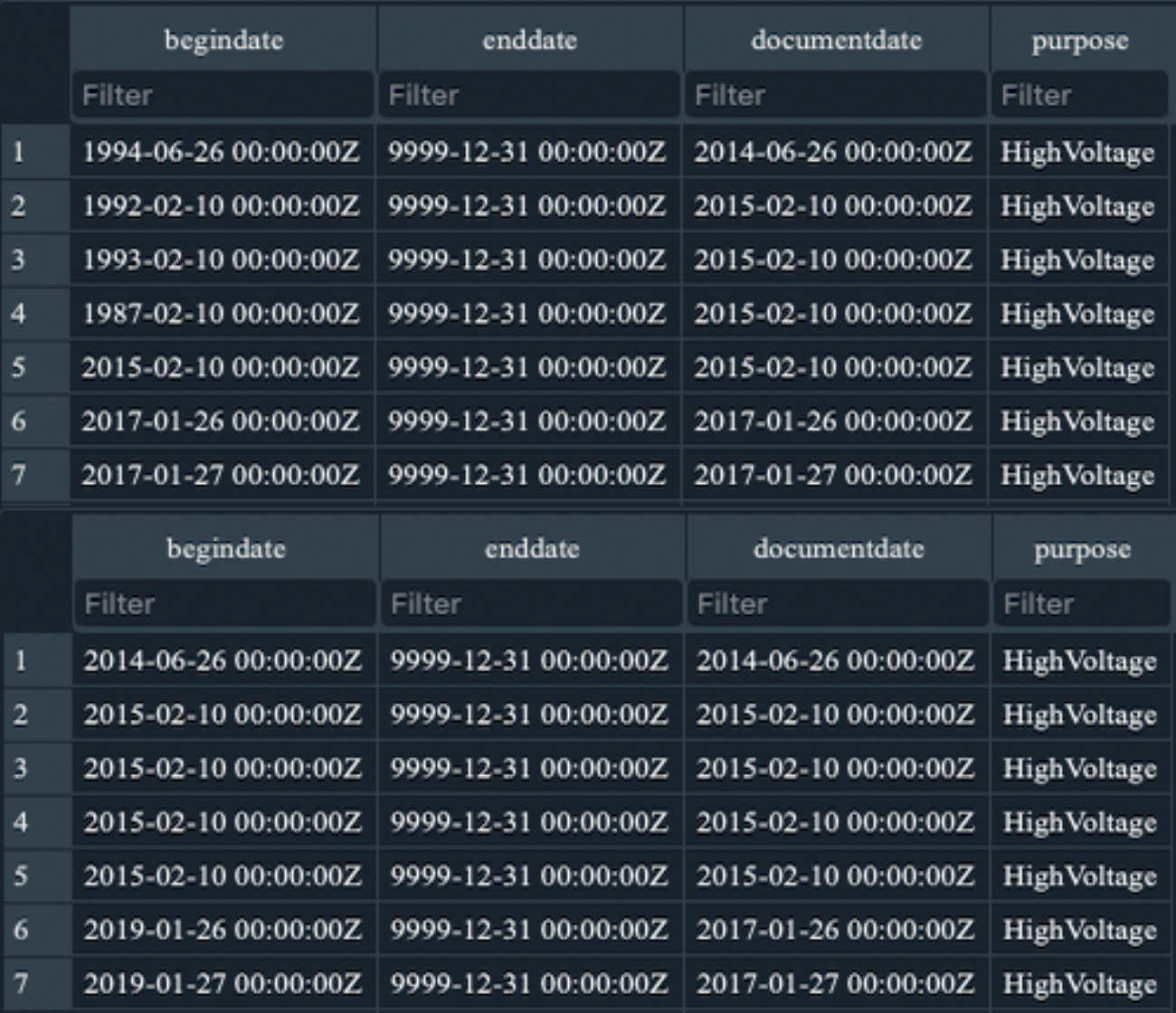}
\caption{Data manipulation in scenario A: the above is the ground truth data, and the below is the modified data.}
\label{scenarioA}
\end{figure}

\begin{figure}[!t]
\centering
\includegraphics[width=0.45\textwidth]{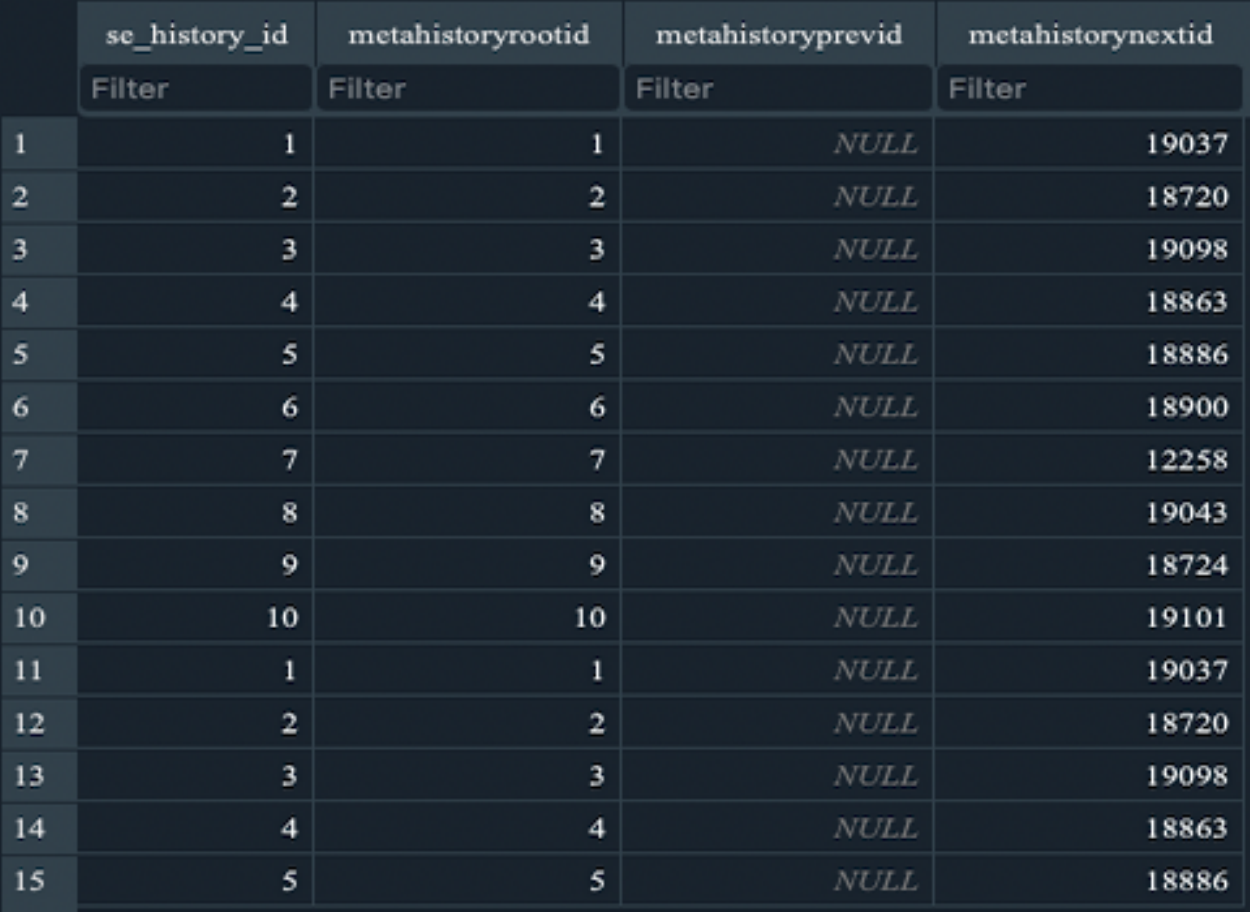}
\caption{Data manipulation in scenario B: the first 10 records are repeated.}
\label{scenarioB}
\end{figure}

\begin{itemize}
\item \emph{Scenario A}: 
Electron, an electricity company, has some bad times due to disappointing sales and aged assets that needed to be replaced. Bob, the CEO of Electron, is afraid that the presentation of this bad news will influence the stock price. To handle this, he asks one of the IT interns, Alice, to make some of the cable data ``newer'' by changing the date they went into the ground. This will increase the apparent value of the company as the new cable is worth more and presents better results. The changes that Alice made in the data warehouse can be found in \cref{scenarioA}. What Bob does not know is that the company recently started using AUDITEM to validate the data warehouse. 
\item \emph{Scenario B}: 
Alice, an intern at Electron, is working on the efficient storage of spatial data in Electron's data warehouse. She needs a subset of the data and stores it in a separate test environment for her research. However, she forgets to change the output location; as a result, the subset is duplicated within the data warehouse. Because she knows about AUDITEM and does not have permission to delete her mess-up, she decides to re-upload the right batch to AUDITEM so that no one else would notice this mistake. In \cref{scenarioB} the mistakes from Alice can be seen. 
\end{itemize}

\begin{figure}[!b]
\centering
\includegraphics[width=0.4\textwidth]{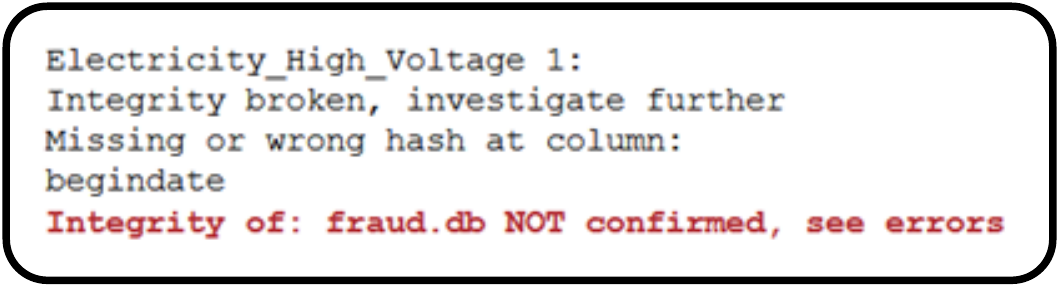}
\caption{Outcome of AUDITEM in Scenario A.}
\label{scenarioResultA}
\end{figure}

\begin{figure}[!t]
\centering
\includegraphics[width=0.5\textwidth]{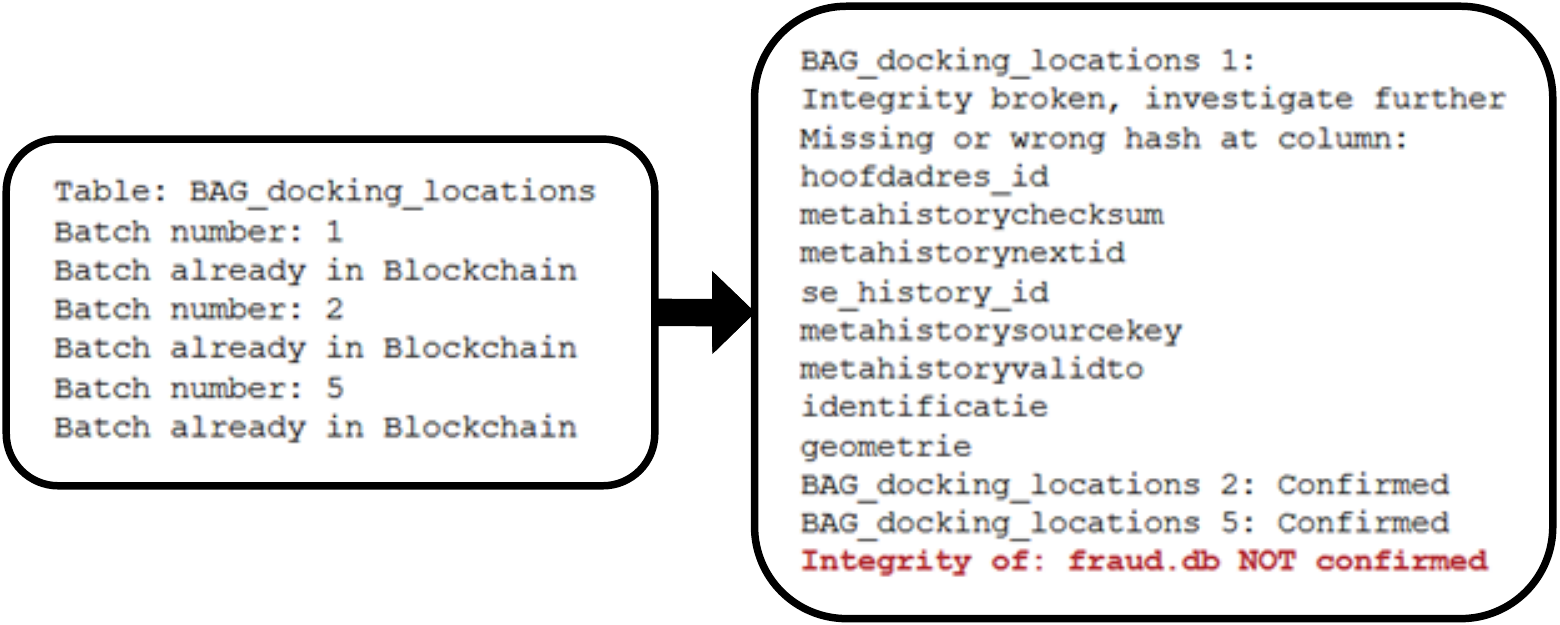}
\caption{Outcome of AUDITEM in scenario B: the first message is on the left, and the second is on the right.}
\label{scenarioResultB}
\end{figure}

\begin{figure*}[!b]
\centering
    \subfloat{
    \centering
    \includegraphics[width=.23\linewidth]{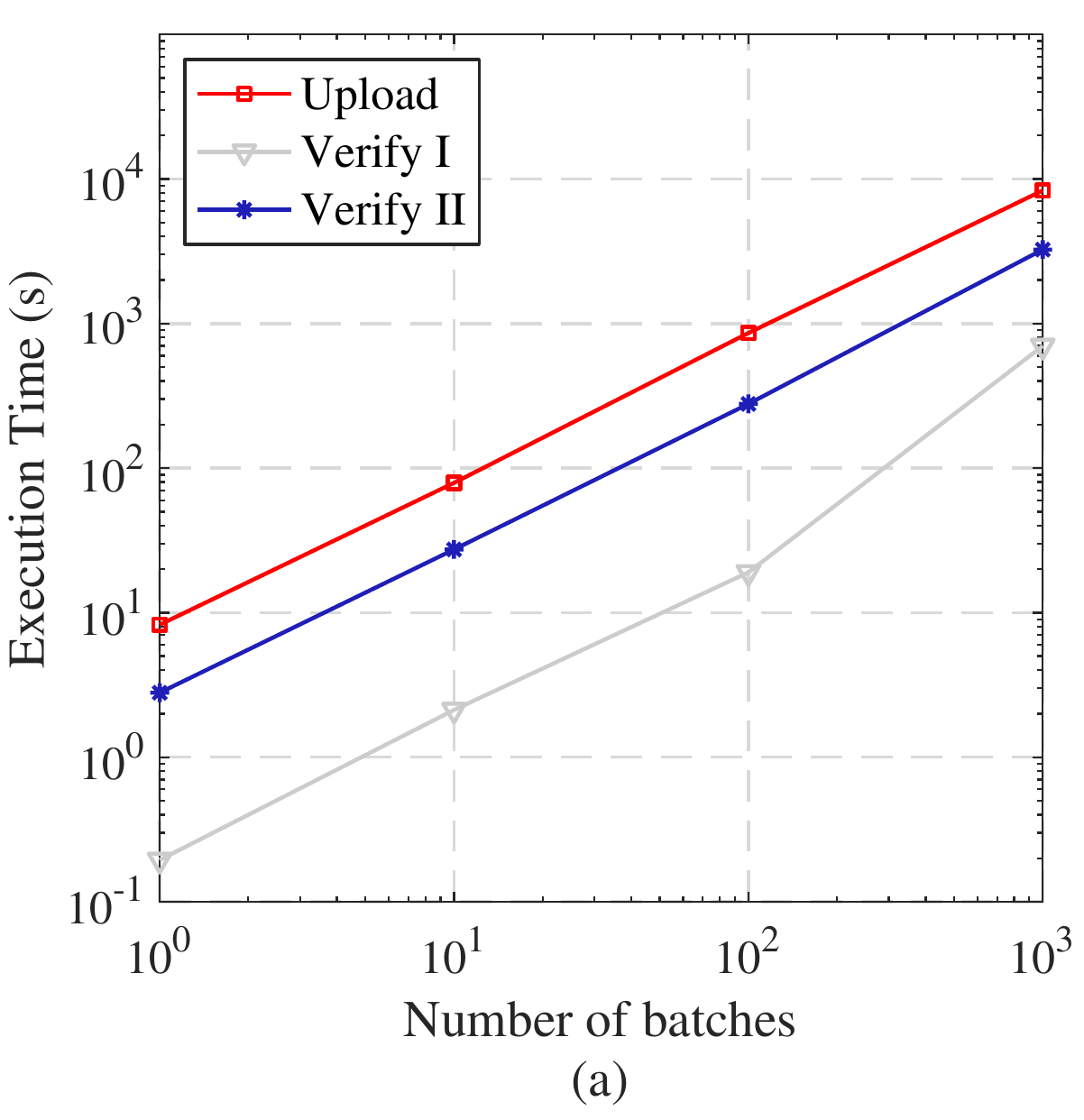}
    \label{api1}
    }\hspace{6em}%
    \subfloat{
    \centering
    \includegraphics[width=.23\linewidth]{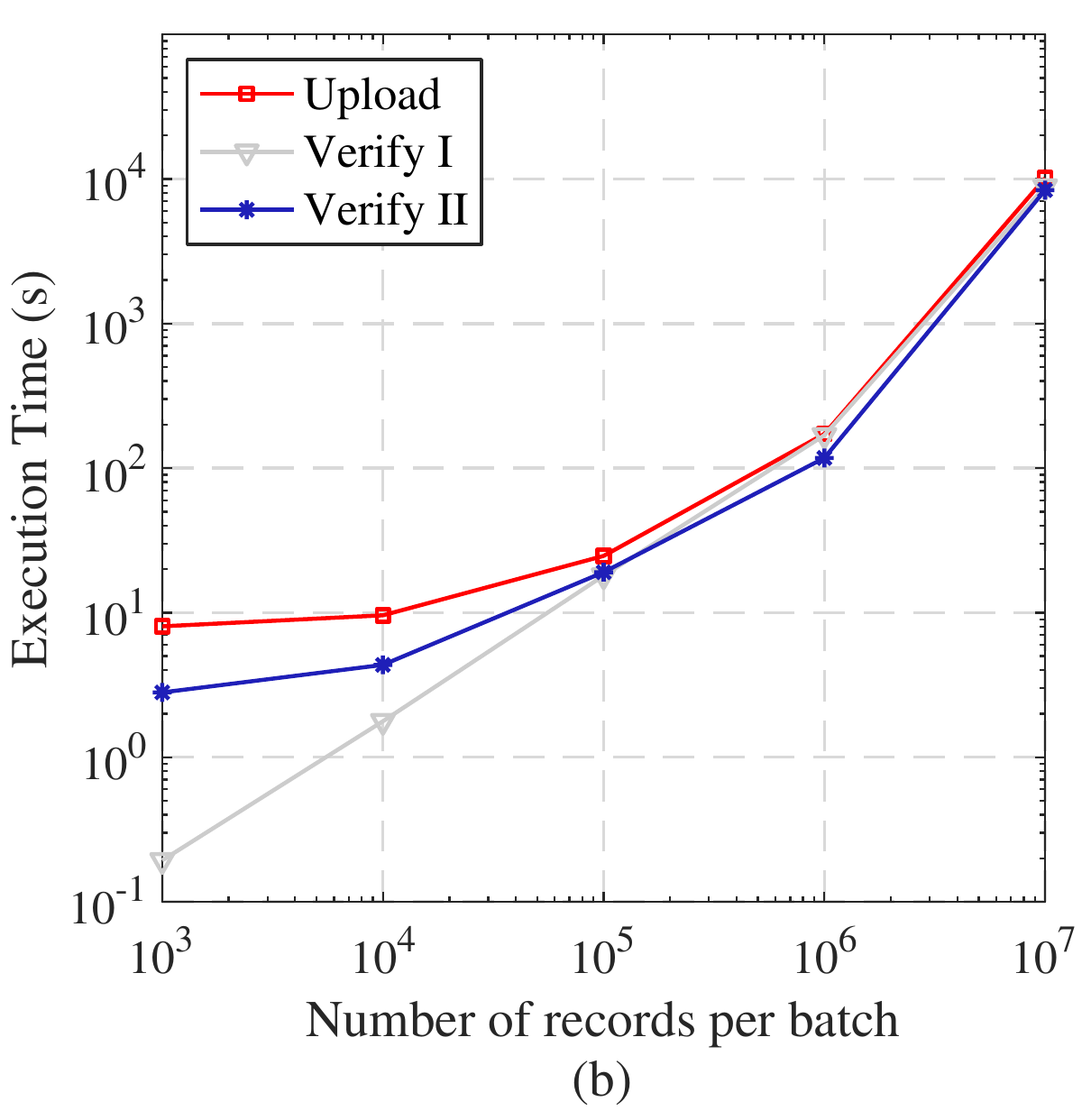}
    \label{api2}
    }\hspace{6em}%
    \subfloat{
    \centering
    \includegraphics[width=.23\linewidth]{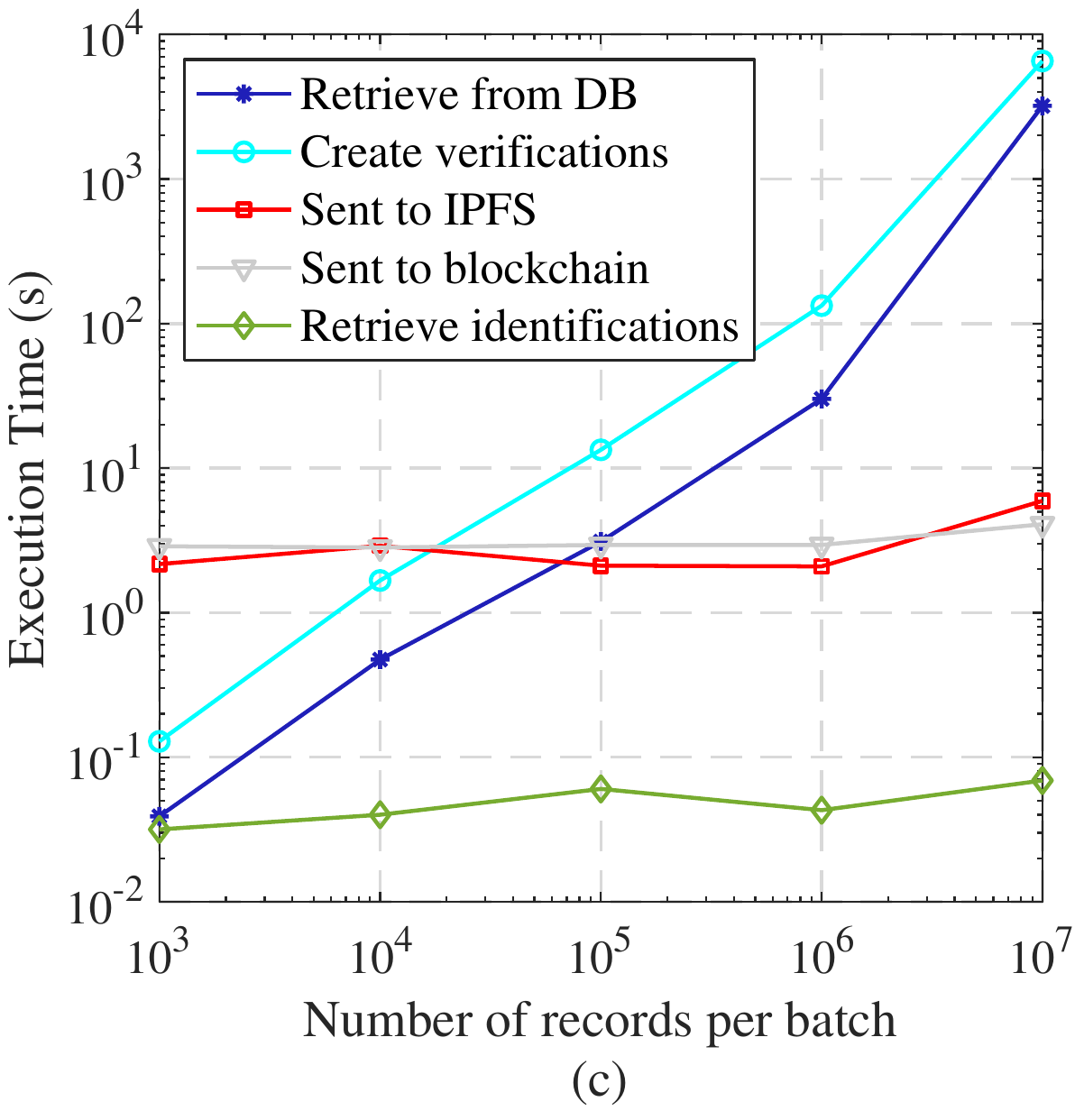}
    \label{api3}
    }\hspace{6em}%
    \caption{Execution time of the DIVT protocol with (a) different batch numbers for upload and verify functions; (b) different batch size for upload and verify functions; and (c) different batch size for individual component functions.}
\label{api}
\end{figure*}

\emph{Results}:
If we reproduce scenario A in AUDITEM, we will get the prompt message presented in \cref{scenarioResultA}. This indicates that when the cable data record is changed, the integrity is reported broken because the hash of the ``begindate'' column is changed. With this result, we can confirm that AUDITEM can detect this kind of fraud, i.e., Alice could not easily manipulate the data, even though she is instructed to do so by the CEO. If we reproduce scenario B in AUDITEM, we have two prompt messages, as shown in \cref{scenarioResultB}. Firstly, the uploading of the new batch is rejected because the batch already exists. 
Secondly, the message shows that all the columns have been changed inside batch 1, but the integrity of non duplicated batches (batch 2 and batch 5) is still correct. This demonstrates that AUDITEM can detect this fraud, i.e., Alice can't hide her mistake by re-uploading the data batch.
It should be noted that nothing prevents Alice from changing the original data in the above scenarios. That is why we designed two verification methods in AUDITEM. By comparing the integrity report with the backups in IPFS, problematic data fields can be identified, and errors can be fixed.

\subsection{Computational Overhead}

To verify whether AUDITEM can handle large and complex data warehouses, we conducted performance tests using multiple test cases. Especially, we repeatedly run AUDITEM with different database settings to find out how long it takes to perform the functions. The database varies in the number of batches and the number of records per batch. For the evaluation, an SQLite database containing information about properties in the Netherlands is chosen to simulate a realistic scenario. A subset table containing 28 columns, including two columns of complex geometries, is used to perform the benchmark. The subsets are generated by randomly selecting 1 000, 10 000, 100 000, 1 000 000, or 10 000 000 records. Then, each subset is duplicated to create 10 subsets. Half of the subsets contain one batch, while the other half contains one batch per 1 000 records. An overview of the subsets can be seen in \cref{batch}. The subset containing 10 000 000 records is only tested once. All other benchmarks are performed in triplicate. For the benchmark, there are three points of interest: 1) the time for uploading subsets to the network, 2) the time for performing integrity check with \emph{Verify I} and \emph{Verify II}; and 3) the time of the individual functions in the DIVT protocol. 

\begin{table}[htb]
\centering
\caption{Overview of variations in benchmark data subsets.}
\resizebox{\columnwidth}{!}{%
\begin{tabular}{@{}lllllll@{}}
\toprule
\multicolumn{2}{c}{Experiment}   & 1 & 2  & 3  & 4   & 5   \\ \midrule

\multirow{2}{*} {Benchmark I} & Records: & $1\,000$ & $10\,000$ & $100\,000$ & $1\,000\,000$ & $10\,000\,000$ \\
        & Batches: & $1$ & $10$  & $100$  & $1\,000$  & -   \\ 
\multirow{2}{*}{Benchmark II} & Records: & $1\,000$ & $10\,000$ & $100\,000$ & $1\,000\,000$ & $10\,000\,000$ \\
        & Batches: & $1$ & $1$  & $1$  & $1$ & $1$   \\ \bottomrule
\end{tabular}
\label{batch}
}
\end{table}

\hyperref[api]{\cref*{api}(a)} shows the overhead variation with different number of batches. 
The result indicates that the execution time to upload one batch is consistent per experiment. As the number of batches increases from 1, 10, 100 to 1 000, the execution time of \emph{Upload} and \emph{Verify II} also increases with a factor of 10. The only exception where the linear increase is not consistent is \emph{Verify I} with 1 000 batches. This higher increase could be explained by a higher blockchain query time.
For the next benchmark, the performance is evaluated with a different number of records per batch, as shown in \hyperref[api]{\cref*{api}(b)}. It can be seen that the execution time of \emph{Verify I} is significantly lower when the batch size is small, but increases to the same level as \emph{Upload} and \emph{Verify II} as the batch size increases. 
Also, it is worth noting that the execution time of the three functions tends to increase exponentially when the number of records increases.
We can conclude that uploading and verifying a batch containing 1 000 000 records is an efficient choice since it takes only 100 seconds.
When the batch size is increased by a further 10 times, the execution time increases by a factor of 100.
Besides, it is not efficient to upload small batches of 1 000 records as this takes almost the same time as 10 000 records.

To investigate this further, the execution time of individual functions with different batch sizes is tested and plotted in \hyperref[api]{\cref*{api}(c)}. It is possible to see which functions cause the system delay in the plot. First, retrieving identifications is the least time-consuming function and remains stable as the number of records increases.
When having 1 000 records per batch, the dominant factors are the two functions that perform API requests and wait for responses: send to IPFS and send to blockchain.
When more records are added to the batch, the execution time of these two functions remains stable. After 100 000 records per batch, retrieving the data from the database and creating the verification attributes are the most dominant factors. This is reasonable since 10 000 000 geographic records are almost 4 GB in size to be transferred to DIVT. In summary, we can conclude that AUDITEM is effective because a large database of 10 000 000 geographical records can be validated within 3 hours.

\subsection{Blockchain Scalability}
\label{scalability}
The limited scalability of blockchain networks often leads to high latency or transaction failures \cite{shi2019operating}. For this reason, AUDITEM aims to limit the number of transactions done with the blockchain. Nevertheless, the blockchain network needs to process transactions from multiple users simultaneously, making a performance evaluation necessary. For the evaluation, a benchmark tool developed by the Hyperledger Foundation called Caliper \cite{caliper}
is used. Caliper is installed as a docker image and run inside a docker container. The performance benchmarks are operated based on the following configurations:

\begin{itemize}
\item In addition to the Hyperledger Fabric mentioned in \cref{implementation}, four other popular permissioned blockchain platforms were identified as benchmarks for testing the AUDITEM system. These platforms include the Hyperledger community's Iroha, Besu, and Sawtooth and the Ethereum permissioned (private) blockchain.
The comparison of these platforms is summarized in \cref{tab:compare}. We leave the testing of more platforms for our future work.
\item The scalability of AUDITEM's underlying blockchain is tested by submitting different numbers of transactions per second (1, 2, 4, 8, 16, 32, 64, 128, 256) to the platform with a fixed workload. Each experiment is repeated five times to obtain the average throughput or latency values.
\item Two basic smart contract operations are measured: 1) Write operations are those related to the creation of attributes required for data integrity verification; 2) Read operations are those related to the query of attributes and certificates in smart contracts.
\item Although different blockchain platforms have different configurations, consensus algorithms, and other factors that may affect performance, we believe these impacts are limited by the platform itself \cite{shi2019operating}. Therefore, we focus on comparing the scalability of these platforms under the benchmark configuration from a platform level.
\item The performance of the blockchain is measured by the metrics of throughput, latency, and success ratio.
In particular, throughput is a measure of how many transactions are completed within a given time frame, latency is the time for transaction execution, and success ratio is the ratio of successful transactions to all submitted transactions \cite{metrics}.
\end{itemize}

\renewcommand{\arraystretch}{0.8}

\begin{table*}[!t]
\small
\centering
\caption{Comparison of five permissioned blockchain platforms.}
\label{tab:compare}
\begin{tabularx}{\textwidth}{m{.1\linewidth}m{.32\linewidth}m{.1\linewidth}m{.12\linewidth}m{.2\linewidth}}
\toprule
\textbf{Blockchain} &
  \textbf{Introduction} &
  \textbf{Consensus Algorithms} &
  \textbf{Smart Contract Protocol} &
  \textbf{Key Features} \\ \midrule
\textbf{Hyperledger Besu} &
  An Ethereum client that can run on public networks, test networks, and private permissioned networks. &
  PoW and PoA &
  Smart Contract &
  \begin{tabular}[c]{@{}l@{}}- Support for multiple types of Ethereum \\ networks\end{tabular} \\
\textbf{Ethereum (private)} &
  An open-source blockchain platform with smart contract functionality that handles transactions via the EVM. &
  PoA &
  Smart Contract &
  \begin{tabular}[c]{@{}l@{}}- Mature blockchain and DApp system\\ - Customized for enterprise applications\end{tabular} \\
\textbf{Hyperledger Fabric} &
  An open-source permissioned blockchain with a highly flexible and adaptable design for enterprise usage. &
  Raft and Kafka (deprecated in v2.x) &
  Chaincode &
  \begin{tabular}[c]{@{}l@{}}- Multi-ledger structure \\ - Private data storage in channels\end{tabular} \\
\textbf{Hyperledger Iroha} &
  A framework for incorporating blockchain into infrastructures or IoT projects that are easy and quick to implement. &
  YAC &
  Command (built in contracts) &
  \begin{tabular}[c]{@{}l@{}}- Easy to use\\  - Built-in smart contract functionality\end{tabular} \\
\textbf{Hyperledger Sawtooth} &
  A secure and modularity-based architecture for creating enterprise-level permissioned blockchains. &
  PoET, Raft, and PBFT &
  Transaction Processor &
  \begin{tabular}[c]{@{}l@{}}- Parallel transaction execution\\ - Customizable transaction processors\end{tabular} \\ \bottomrule
\end{tabularx}%
\end{table*}

\begin{figure*}[!t]
\centering
\includegraphics[width=\textwidth]{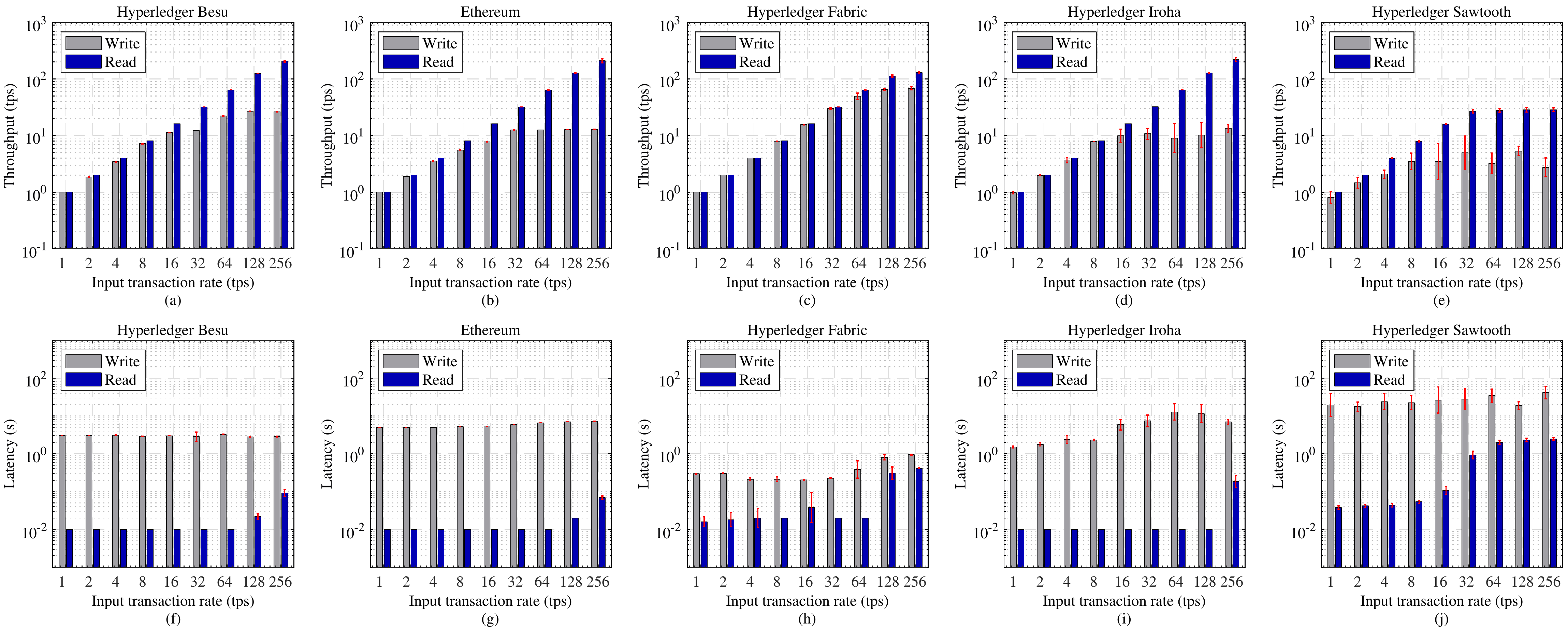}
\caption{Comparison of throughput and latency of write/read operations for five chosen blockchain platforms. The plots (a) to (e) plots show the throughput results, while the plots (f) to (j) show the latency results.}
\label{five}
\end{figure*}

\begin{figure*}[!t]
\centering
\includegraphics[width=\textwidth]{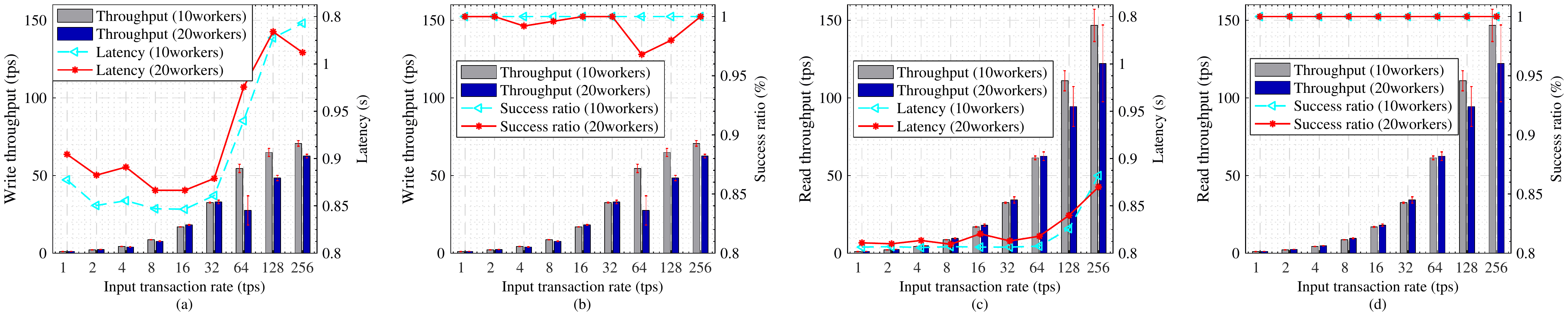}
\caption{Performance of Hyperledger Fabric in the AUDITEM system at different input transaction rates and the number of workers. The plots (a) and (b) show the results of write operations, while the plots (c) and (d) show the results of read operations.}
\label{fabric}
\end{figure*}

\renewcommand{\arraystretch}{0.7}
\begin{table*}[!t]
\centering
\caption{Execution latency of five blockchain platforms.}
\label{tab:latency}
\resizebox{\textwidth}{!}{%
\begin{tabular}{clllllll}
\toprule
\multirow{2}{*}{\textbf{Blockchain}} &
  \multicolumn{1}{c}{\multirow{2}{*}{\textbf{Send Rate (TPS)}}} &
  \multicolumn{3}{c}{\textbf{Write Operation}} &
  \multicolumn{3}{c}{\textbf{Read Operation}} \\ \cline{3-8} 
 &
  \multicolumn{1}{c}{} &
  \multicolumn{1}{c}{\textbf{Max Latency (s)}} &
  \multicolumn{1}{c}{\textbf{Min Latency (s)}} &
  \multicolumn{1}{c}{\textbf{Avg Latency (s)}} &
  \multicolumn{1}{c}{\textbf{Max Latency (s)}} &
  \multicolumn{1}{c}{\textbf{Min Latency (s)}} &
  \multicolumn{1}{c}{\textbf{Avg Latency (s)}} \\ \midrule
\multirow{9}{*}{\textbf{Hyperledger Beau}}     & 1   & 5.116 $\pm$ 0.009   & 1.036 $\pm$ 0.005   & 3.056 $\pm$ 0.005   & 0.032 $\pm$ 0.011 & 0.010 $\pm$ 0.000 & 0.010 $\pm$ 0.000 \\
                                             & 2   & 5.288 $\pm$ 0.443   & 1.030 $\pm$ 0.000   & 3.056 $\pm$ 0.031   & 0.032 $\pm$ 0.022 & 0.010 $\pm$ 0.000 & 0.010 $\pm$ 0.000 \\
                                             & 4   & 5.476 $\pm$ 0.533   & 1.030 $\pm$ 0.000   & 3.108 $\pm$ 0.091   & 0.022 $\pm$ 0.004 & 0.010 $\pm$ 0.000 & 0.010 $\pm$ 0.000 \\
                                             & 8   & 5.656 $\pm$ 0.540   & 1.028 $\pm$ 0.004   & 2.912 $\pm$ 0.051   & 0.024 $\pm$ 0.005 & 0.010 $\pm$ 0.000 & 0.010 $\pm$ 0.000 \\
                                             & 16  & 5.860 $\pm$ 0.447   & 1.028 $\pm$ 0.004   & 3.022 $\pm$ 0.013   & 0.020 $\pm$ 0.000 & 0.010 $\pm$ 0.000 & 0.010 $\pm$ 0.000 \\
                                             & 32  & 5.402 $\pm$ 0.608   & 1.226 $\pm$ 0.449   & 2.850 $\pm$ 0.788   & 0.020 $\pm$ 0.000 & 0.010 $\pm$ 0.000 & 0.010 $\pm$ 0.000 \\
                                             & 64  & 4.082 $\pm$ 0.008   & 2.046 $\pm$ 0.009   & 3.266 $\pm$ 0.045   & 0.026 $\pm$ 0.005 & 0.010 $\pm$ 0.000 & 0.010 $\pm$ 0.000 \\
                                             & 128 & 3.340 $\pm$ 0.119   & 2.154 $\pm$ 0.133   & 2.796 $\pm$ 0.044   & 0.050 $\pm$ 0.029 & 0.010 $\pm$ 0.000 & 0.022 $\pm$ 0.004 \\
                                             & 256 & 3.336 $\pm$ 0.088   & 2.098 $\pm$ 0.035   & 2.844 $\pm$ 0.068   & 0.190 $\pm$ 0.030 & 0.016 $\pm$ 0.005 & 0.092 $\pm$ 0.019 \\ \midrule
\multirow{9}{*}{\textbf{Ethereum (private)}}             & 1   & 7.072 $\pm$ 0.015   & 3.020 $\pm$ 0.000   & 5.034 $\pm$ 0.005   & 0.020 $\pm$ 0.000 & 0.010 $\pm$ 0.000 & 0.010 $\pm$ 0.000 \\
                                             & 2   & 7.070 $\pm$ 0.012   & 3.020 $\pm$ 0.000   & 5.032 $\pm$ 0.004   & 0.020 $\pm$ 0.000 & 0.010 $\pm$ 0.000 & 0.010 $\pm$ 0.000 \\
                                             & 4   & 7.056 $\pm$ 0.009   & 3.020 $\pm$ 0.000   & 5.030 $\pm$ 0.000   & 0.020 $\pm$ 0.000 & 0.010 $\pm$ 0.000 & 0.010 $\pm$ 0.000 \\
                                             & 8   & 7.060 $\pm$ 0.000   & 2.822 $\pm$ 0.443   & 5.250 $\pm$ 0.022   & 0.022 $\pm$ 0.004 & 0.010 $\pm$ 0.000 & 0.010 $\pm$ 0.000 \\
                                             & 16  & 7.074 $\pm$ 0.005   & 3.024 $\pm$ 0.005   & 5.390 $\pm$ 0.025   & 0.030 $\pm$ 0.022 & 0.010 $\pm$ 0.000 & 0.010 $\pm$ 0.000 \\
                                             & 32  & 7.062 $\pm$ 0.004   & 4.034 $\pm$ 0.005   & 5.920 $\pm$ 0.019   & 0.020 $\pm$ 0.000 & 0.010 $\pm$ 0.000 & 0.010 $\pm$ 0.000 \\
                                             & 64  & 7.086 $\pm$ 0.005   & 6.022 $\pm$ 0.004   & 6.624 $\pm$ 0.005   & 0.028 $\pm$ 0.008 & 0.010 $\pm$ 0.000 & 0.010 $\pm$ 0.000 \\
                                             & 128 & 7.184 $\pm$ 0.055   & 7.036 $\pm$ 0.005   & 7.084 $\pm$ 0.017   & 0.038 $\pm$ 0.008 & 0.010 $\pm$ 0.000 & 0.020 $\pm$ 0.000 \\
                                             & 256 & 7.468 $\pm$ 0.182   & 4.060 $\pm$ 2.748   & 7.258 $\pm$ 0.112   & 0.162 $\pm$ 0.025 & 0.012 $\pm$ 0.004 & 0.070 $\pm$ 0.007 \\ \midrule
\multirow{9}{*}{\textbf{Hyperledger Fabric}}   & 1   & 0.360 $\pm$ 0.101   & 0.288 $\pm$ 0.004   & 0.298 $\pm$ 0.008   & 0.022 $\pm$ 0.004 & 0.010 $\pm$ 0.000 & 0.016 $\pm$ 0.005 \\
                                             & 2   & 0.694 $\pm$ 0.518   & 0.274 $\pm$ 0.036   & 0.304 $\pm$ 0.005   & 0.192 $\pm$ 0.373 & 0.010 $\pm$ 0.000 & 0.018 $\pm$ 0.008 \\
                                             & 4   & 0.518 $\pm$ 0.448   & 0.040 $\pm$ 0.000   & 0.214 $\pm$ 0.017   & 0.206 $\pm$ 0.416 & 0.010 $\pm$ 0.000 & 0.020 $\pm$ 0.012 \\
                                             & 8   & 0.528 $\pm$ 0.454   & 0.048 $\pm$ 0.004   & 0.214 $\pm$ 0.031   & 0.028 $\pm$ 0.018 & 0.010 $\pm$ 0.000 & 0.020 $\pm$ 0.000 \\
                                             & 16  & 0.342 $\pm$ 0.018   & 0.052 $\pm$ 0.004   & 0.204 $\pm$ 0.005   & 0.224 $\pm$ 0.456 & 0.010 $\pm$ 0.000 & 0.038 $\pm$ 0.040 \\
                                             & 32  & 0.388 $\pm$ 0.029   & 0.072 $\pm$ 0.011   & 0.226 $\pm$ 0.005   & 0.036 $\pm$ 0.005 & 0.010 $\pm$ 0.000 & 0.020 $\pm$ 0.000 \\
                                             & 64  & 0.726 $\pm$ 0.446   & 0.132 $\pm$ 0.029   & 0.386 $\pm$ 0.214   & 0.052 $\pm$ 0.022 & 0.010 $\pm$ 0.000 & 0.020 $\pm$ 0.000 \\
                                             & 128 & 1.164 $\pm$ 0.153   & 0.324 $\pm$ 0.113   & 0.810 $\pm$ 0.132   & 0.698 $\pm$ 0.178 & 0.026 $\pm$ 0.009 & 0.308 $\pm$ 0.120 \\
                                             & 256 & 1.238 $\pm$ 0.099   & 0.532 $\pm$ 0.023   & 0.946 $\pm$ 0.038   & 0.760 $\pm$ 0.034 & 0.042 $\pm$ 0.004 & 0.416 $\pm$ 0.011 \\ \midrule
\multirow{9}{*}{\textbf{Hyperledger Iroha}}    & 1   & 5.202 $\pm$ 1.894   & 0.106 $\pm$ 0.015   & 1.518 $\pm$ 0.078   & 0.024 $\pm$ 0.005 & 0.010 $\pm$ 0.000 & 0.010 $\pm$ 0.000 \\
                                             & 2   & 5.760 $\pm$ 0.372   & 0.168 $\pm$ 0.018   & 1.786 $\pm$ 0.174   & 0.018 $\pm$ 0.004 & 0.010 $\pm$ 0.000 & 0.010 $\pm$ 0.000 \\
                                             & 4   & 8.602 $\pm$ 2.327   & 0.164 $\pm$ 0.011   & 2.394 $\pm$ 0.573   & 0.018 $\pm$ 0.004 & 0.010 $\pm$ 0.000 & 0.010 $\pm$ 0.000 \\
                                             & 8   & 6.350 $\pm$ 0.587   & 0.166 $\pm$ 0.054   & 2.346 $\pm$ 0.111   & 0.026 $\pm$ 0.019 & 0.010 $\pm$ 0.000 & 0.010 $\pm$ 0.000 \\
                                             & 16  & 9.746 $\pm$ 3.528   & 2.256 $\pm$ 0.942   & 5.890 $\pm$ 1.966   & 0.018 $\pm$ 0.004 & 0.010 $\pm$ 0.000 & 0.010 $\pm$ 0.000 \\
                                             & 32  & 9.188 $\pm$ 3.140   & 4.506 $\pm$ 1.388   & 7.416 $\pm$ 2.729   & 0.020 $\pm$ 0.000 & 0.010 $\pm$ 0.000 & 0.010 $\pm$ 0.000 \\
                                             & 64  & 15.542 $\pm$ 9.606  & 10.172 $\pm$ 4.662  & 12.906 $\pm$ 6.709  & 0.020 $\pm$ 0.012 & 0.010 $\pm$ 0.000 & 0.010 $\pm$ 0.000 \\
                                             & 128 & 14.048 $\pm$ 10.550 & 9.152 $\pm$ 4.815   & 11.530 $\pm$ 6.690  & 0.026 $\pm$ 0.005 & 0.010 $\pm$ 0.000 & 0.010 $\pm$ 0.000 \\
                                             & 256 & 7.532 $\pm$ 1.267   & 5.618 $\pm$ 0.224   & 7.018 $\pm$ 1.116   & 0.378 $\pm$ 0.107 & 0.016 $\pm$ 0.005 & 0.186 $\pm$ 0.069 \\ \midrule
\multirow{9}{*}{\textbf{Hyperledger Sawtooth}} & 1 & 56.242 $\pm$ 30.050 & 0.552 $\pm$ 0.180 & 19.552 $\pm$ 14.957 & 0.062 $\pm$ 0.013 & 0.030 $\pm$ 0.000 & 0.038 $\pm$ 0.004 \\
                                             & 2   & 38.082 $\pm$ 11.623 & 1.124 $\pm$ 0.475   & 17.746 $\pm$ 4.899  & 0.456 $\pm$ 0.189 & 0.030 $\pm$ 0.000 & 0.042 $\pm$ 0.004 \\
                                             & 4   & 35.822 $\pm$ 10.701 & 13.658 $\pm$ 11.498 & 23.844 $\pm$ 12.153 & 0.454 $\pm$ 0.075 & 0.030 $\pm$ 0.000 & 0.044 $\pm$ 0.005 \\
                                             & 8   & 28.694 $\pm$ 9.826  & 3.676 $\pm$ 6.420   & 22.568 $\pm$ 9.709  & 0.570 $\pm$ 0.099 & 0.030 $\pm$ 0.000 & 0.054 $\pm$ 0.005 \\
                                             & 16  & 29.628 $\pm$ 23.472 & 23.430 $\pm$ 23.532 & 26.646 $\pm$ 23.373 & 0.680 $\pm$ 0.137 & 0.030 $\pm$ 0.000 & 0.108 $\pm$ 0.028 \\
                                             & 32  & 29.944 $\pm$ 18.929 & 26.612 $\pm$ 19.225 & 28.322 $\pm$ 19.015 & 1.270 $\pm$ 0.249 & 0.050 $\pm$ 0.012 & 0.926 $\pm$ 0.219 \\
                                             & 64  & 35.388 $\pm$ 14.294 & 24.278 $\pm$ 15.852 & 34.504 $\pm$ 14.149 & 2.672 $\pm$ 0.227 & 0.066 $\pm$ 0.015 & 2.010 $\pm$ 0.244 \\
                                             & 128 & 19.414 $\pm$ 4.372  & 15.428 $\pm$ 7.824  & 18.946 $\pm$ 4.363  & 2.860 $\pm$ 0.321 & 0.188 $\pm$ 0.097 & 2.320 $\pm$ 0.263 \\
                                             & 256 & 41.724 $\pm$ 15.870 & 41.260 $\pm$ 15.857 & 41.486 $\pm$ 15.867 & 3.062 $\pm$ 0.263 & 0.634 $\pm$ 0.071 & 2.504 $\pm$ 0.194 \\ \bottomrule  
\end{tabular}%
}
\end{table*}

\renewcommand{\arraystretch}{0.7}
\begin{table*}[!t]
\centering
\caption{Execution latency of Hyperledger Fabric with different worker numbers.}
\label{tab:fabriclatency}
\resizebox{\textwidth}{!}{%
\begin{tabular}{clllllll}
\toprule
\multirow{2}{*}{\textbf{Blockchain}} &
  \multicolumn{1}{c}{\multirow{2}{*}{\textbf{Send Rate (TPS)}}} &
  \multicolumn{3}{c}{\textbf{Write Operation}} &
  \multicolumn{3}{c}{\textbf{Read Operation}} \\ \cline{3-8} 
 &
  \multicolumn{1}{c}{} &
  \multicolumn{1}{c}{\textbf{Max Latency (s)}} &
  \multicolumn{1}{c}{\textbf{Min Latency (s)}} &
  \multicolumn{1}{c}{\textbf{Avg Latency (s)}} &
  \multicolumn{1}{c}{\textbf{Max Latency (s)}} &
  \multicolumn{1}{c}{\textbf{Min Latency (s)}} &
  \multicolumn{1}{c}{\textbf{Avg Latency (s)}} \\ \midrule

\multirow{9}{*}{\shortstack[c]{\textbf{Hyperledger Fabric} \\ \textbf{(10 workers)}}} 
                                                 & 1   & 0.684 $\pm$ 0.390     & 0.124 $\pm$ 0.023     & 0.310 $\pm$ 0.057     & 0.074 $\pm$ 0.022     & 0.010 $\pm$ 0.000     & 0.026 $\pm$ 0.005     \\
                                                 & 2   & 0.342 $\pm$ 0.049     & 0.070 $\pm$ 0.000     & 0.202 $\pm$ 0.013     & 0.086 $\pm$ 0.030     & 0.012 $\pm$ 0.004     & 0.028 $\pm$ 0.008     \\
                                                 & 4   & 0.370 $\pm$ 0.025     & 0.082 $\pm$ 0.013     & 0.222 $\pm$ 0.026     & 0.056 $\pm$ 0.009     & 0.010 $\pm$ 0.000     & 0.024 $\pm$ 0.005     \\
                                                 & 8   & 0.364 $\pm$ 0.130     & 0.070 $\pm$ 0.007     & 0.188 $\pm$ 0.022     & 0.074 $\pm$ 0.018     & 0.010 $\pm$ 0.000     & 0.028 $\pm$ 0.004     \\
                                                 & 16  & 0.336 $\pm$ 0.044     & 0.076 $\pm$ 0.013     & 0.186 $\pm$ 0.009     & 0.062 $\pm$ 0.023     & 0.010 $\pm$ 0.000     & 0.026 $\pm$ 0.005     \\
                                                 & 32  & 0.558 $\pm$ 0.566     & 0.070 $\pm$ 0.010     & 0.244 $\pm$ 0.160     & 0.126 $\pm$ 0.128     & 0.012 $\pm$ 0.004     & 0.026 $\pm$ 0.013     \\
                                                 & 64  & 0.802 $\pm$ 0.130     & 0.216 $\pm$ 0.057     & 0.560 $\pm$ 0.079     & 0.108 $\pm$ 0.024     & 0.010 $\pm$ 0.000     & 0.030 $\pm$ 0.000     \\
                                                 & 128 & 1.206 $\pm$ 0.042     & 0.610 $\pm$ 0.046     & 0.910 $\pm$ 0.032     & 0.274 $\pm$ 0.032     & 0.012 $\pm$ 0.004     & 0.104 $\pm$ 0.015     \\
                                                 & 256 & 1.236 $\pm$ 0.009     & 0.678 $\pm$ 0.049     & 0.972 $\pm$ 0.018     & 0.548 $\pm$ 0.031     & 0.012 $\pm$ 0.004     & 0.328 $\pm$ 0.019     \\ \midrule
\multirow{9}{*}{\shortstack[c]{\textbf{Hyperledger Fabric} \\ \textbf{(20 workers)}}} 
                                                 & 1   & 0.892 $\pm$ 0.442     & 0.136 $\pm$ 0.061     & 0.418 $\pm$ 0.104     & 0.134 $\pm$ 0.054     & 0.012 $\pm$ 0.004     & 0.044 $\pm$ 0.026     \\
                                                 & 2   & 0.604 $\pm$ 0.090     & 0.120 $\pm$ 0.026     & 0.330 $\pm$ 0.033     & 0.144 $\pm$ 0.092     & 0.014 $\pm$ 0.005     & 0.040 $\pm$ 0.014      \\
                                                 & 4   & 1.430 $\pm$ 1.743     & 0.133 $\pm$ 0.036     & 0.333 $\pm$ 0.043     & 0.200 $\pm$ 0.175     & 0.012 $\pm$ 0.004     & 0.054 $\pm$ 0.043     \\
                                                 & 8   & 0.485 $\pm$ 0.062     & 0.108 $\pm$ 0.025     & 0.265 $\pm$ 0.026     & 0.100 $\pm$ 0.031     & 0.014 $\pm$ 0.005     & 0.038 $\pm$ 0.004     \\
                                                 & 16  & 0.444 $\pm$ 0.055     & 0.124 $\pm$ 0.034     & 0.266 $\pm$ 0.021     & 0.248 $\pm$ 0.224     & 0.016 $\pm$ 0.009     & 0.082 $\pm$ 0.105     \\
                                                 & 32  & 0.594 $\pm$ 0.203     & 0.120 $\pm$ 0.027     & 0.316 $\pm$ 0.086     & 0.124 $\pm$ 0.106     & 0.012 $\pm$ 0.004     & 0.052 $\pm$ 0.050     \\
                                                 & 64  & 0.860 $\pm$ 0.121     & 0.230 $\pm$ 0.092     & 0.547 $\pm$ 0.111     & 0.264 $\pm$ 0.328     & 0.012 $\pm$ 0.004     & 0.072 $\pm$ 0.069     \\
                                                 & 128 & 1.175 $\pm$ 0.114     & 0.683 $\pm$ 0.047     & 0.945 $\pm$ 0.068     & 0.320 $\pm$ 0.289     & 0.020 $\pm$ 0.007     & 0.160 $\pm$ 0.174      \\
                                                 & 256 & 1.390 $\pm$ 0.054     & 0.818 $\pm$ 0.061     & 1.060 $\pm$ 0.029     & 0.490 $\pm$ 0.047     & 0.018 $\pm$ 0.004     & 0.280 $\pm$ 0.029   \\ \bottomrule  
\end{tabular}%
}
\end{table*}

\cref{five} contains ten plots comparing the five blockchain platforms' throughput and latency for write/read operations with 1 worker. First, Beau, Ethereum, and Iroha have similar performance in the read throughput testing, with a bottleneck of around 200 transactions per second (TPS). Fabric holds a slightly lower read throughput of around 130 TPS. Sawtooth offers the worst performance, with a bottleneck of only 30 TPS. In terms of write performance, Ethereum and Iroha have similar write throughput bottlenecks of about 13 TPS. In contrast, Besu has about twice as much throughput when input transaction rates are very high. Fabric holds a much higher write throughput than other platforms, with a bottleneck of around 70 TPS. Secondly, the five subplots below show that Fabric has significantly lower latency for write operations than the other four blockchain platforms. A detailed comparison of the maximum, minimum, and average latency of the five blockchain platforms can be seen in \cref{tab:latency}. The result further validates our conclusion that although Fabric's read performance (throughput and latency) is not the best, it has the best performance for write operations and acceptable performance for read operations. In fact, write performance is particularly critical in our model because more transactions submitted to the blockchain imply a higher level of security for integrity verification. In contrast, read performance is not as critical since data auditing is not required at high frequency. Based on this experiment, we conclude that Fabric is the blockchain platform that best meets our scalability requirement. 

\cref{fabric} further interprets the performance of the Fabric under different worker numbers. At this time, 
different numbers of workers (10 and 20) are leveraged to diversify clients connected to the blockchain network. 
As can be seen from the plot, the highest write/read throughput can be found for both numbers of workers at an input transaction rate of 256 TPS. When the number of workers is doubled from 10 to 20, the measurement's throughput and stability are slightly decreased. However, the overall performance bottleneck is maintained at a high level (write and read throughput are 60 TPS and 120 TPS for 20 workers). The performance is also compared with the transaction success ratio, as shown in \hyperref[fabric]{\cref*{fabric}(b)} and \hyperref[fabric]{\cref*{fabric}(d)}. It can be seen that when the number of workers is 10, all transactions can be submitted successfully. However, when worker numbers increase to 20, some of the write operations may fail when the input transaction rate is high. Nevertheless, the success ratio is maintained above 95\% in all experiments, which demonstrates the good scalability of the model. It is also evident from the latency experiments in \hyperref[fabric]{\cref*{fabric}(a)} and \hyperref[fabric]{\cref*{fabric}(c)} that the latency increases 
in different degrees when both the number of workers and the input transaction rate increase. However, all the latency values are maintained within 1.5 seconds. 
This low latency of the Fabric blockchain ensures that the entire model can work efficiently.

In \cref{tab:fabriclatency}, we show the latency result of Fabric with different worker numbers. It can be seen that despite some fluctuations, the maximum, minimum, and average latency increases for both worker numbers when increasing the input transaction rate from 1 to 256 TPS. If we take the example of 1 000 verification transactions, in the worst case (i.e., an input rate of 256), 10 and 20 workers can still complete the workload in an average of 16 and 18 minutes. Latency only increased by 9\% when the number of workers doubled. Compared to write operations, read operations have very low latency, with 1 000 queries costing only about 30 seconds to complete.
These experiments demonstrate that the latency satisfies the scalability requirements of our model.

\section{Discussion}
\label{discussion}
This section discusses the advantages and limitations of the proposed model.

\subsection{Advantages} 
We first discuss the advantages of AUDITEM based on the research questions proposed in \cref{rqs}.

\newtheorem{thm}{RQ}
\begin{thm}
What blockchain technologies are suitable for data verification, and how scalable are these technologies when more data need to be validated?
\end{thm}

We have identified that scalability is one of the core challenges of combining blockchain with large-scale data verification tasks. It is well known that public blockchain networks such as Bitcoin and Ethereum suffer from low throughput and high transaction costs. Therefore, a permissioned blockchain with better scalability and privacy properties is decided as the underlying blockchain of AUDITEM for data verification. It should be noted that the scalability issues may also arise in permissioned blockchains, which is why we conducted various experiments in \cref{scalability}. The experimental results demonstrated that Hyperledger Fabric is the one that best meets our scalability requirement compared to the other four benchmark blockchain platforms. Besides, AUDITEM tries to minimize the transactions submitted to the blockchain, as this is often the most limiting factor for scalability. A distributed file system is used in AUDITEM to store so-called verification attributes to validate or recover data repositories. These attributes allow flexibility in choosing the degree of recoverability and traceability when more data need to be validated. 

\begin{thm}
How to prevent blockchain from sharing private company data with unauthorized parties while still ensuring data integrity?
\end{thm}

In AUDITEM, a series of approaches are taken to prevent private data from being shared with unauthorized parties. These methods include the careful selection of what is stored on the blockchain, the encryption of the verification attributes, and the use of a private storage chaincode to store private keys. In order to protect private data, only non-personally identification attributes are stored on the blockchain in AUDITEM. In addition, to ensure privacy and traceability, AUDITEM utilizes both hashing and AES encryption to prevent unauthorized access to the company's private data. Hashing the data table identifies the location of possible tampering while protecting data privacy very well. All verification records are encrypted using AES and stored in a decentralized file system for further validation. Finally, we take advantage of Hyperledger Fabric's ability to create private data collections and use a private storage chaincode to generate and query private keys. These private keys are stored on private channels in Hyperledger Fabric and are only accessible to authorized organizations/peers. With this chaincode, private keys can be securely shared with external auditors when needed.

\begin{thm}
How to comply with national personal data regulations when personal data is stored in blockchain for data verification?
\end{thm}

One of the core features of blockchain is the immutability of the data. This is by nature conflicting with data regulations, e.g., the right to be forgotten in GDPR. Therefore, a solution is not as straightforward as expected. 
Storing personal data or references in immutable integrity files gives them a trustworthy status. If AUDITEM allows integrity files to be updated when personal data is changed or deleted, the malicious actor can use this function for their benefit. To solve this, we designed an update function that can only be invoked when personal data is edited. The smart contract confirms whether only the personal data is edited and the non-personal data is authentic before updating the data. The user and the reason for calling the function are kept as logs in the blockchain for future auditing. With this solution, personal data can be updated when needed, but it is difficult to change non-personal data and perform malicious actions. For further research, it would be beneficial to implement AUTITEM and confirm its compatibility with other data regulations.

Next, we discuss how AUDITEM fills the research gaps summarized in \cref{gaps}. 
AUDITEM is compatible with most existing data management tools without changing the core system. For regular users, the activity of adding or querying data within the data repository does not change. Besides, the workflow is totally automated and optimized for data verification processes; users and internal/external auditors have a new interface to upload verification attributes and check the data integrity. To validate how AUDITEM works with a complex data type, the geometric data has been tested in \cref{evaluation} with good performance results obtained.
Secondly, AUDITEM is decentralized and does not require a trusted third party to perform integrity verification. It leverages a blockchain to provide immutable and trustworthy certificates and an IPFS to assist the storage of additional data. All verification tasks are performed automatically based on self-executing smart contracts and the DIVT protocol.
Regarding scalability, AUDITEM works with batch validation to reduce costs. This allows validating a database on-demand at low overhead. In addition, AUDITEM is developed based on the permissioned blockchain, and experiments have shown that the current implementation of Hyperledger Fabric has good scalability.
Finally, companies expect to update or delete personal data in a timely manner. AUDITEM does not change the current way GDPR is handled. It still gives the possibility to delete or edit personal information without altering the blockchain. 
So companies that are currently compliant with the GDPR can expect to remain compliant with the GDPR after implementing AUDITEM.

\subsection{Limitations} 
The limitations of AUDITEM can be discussed in terms of scalability and recoverability concerns and security issues.

\subsubsection{Scalability and Recoverability Concerns}
We observed that when performing upload and verification operations, it is not recommended to use data batches with a large number of records because of the long processing time. A solution to this problem could be concurrent processing, where multiple small data batches are uploaded simultaneously; however, this also increases the transaction input rate to the blockchain and can lead to high latency. In the current implementation of AUDITEM, we observed that the Hyperledger Fabric blockchain could not handle write operations with more than 70 TPS. This can lead to problems when companies upload their initial database or use cases where uploads are more frequent. 
In this regard, more advanced blockchain scalability solutions such as sharding and DAG blockchains \cite{li2020survey} should be investigated in the future.

The next concern is the degree of traceability of malicious behaviors and the recoverability of tampered data, as AUDITEM does not prevent tampering. Storing more data in integrity files increases traceability and recoverability; this does not increase the workload of the blockchain but does result in increased storage in the IPFS. The question of how IPFS handles large-scale file storage has not been investigated. To completely prevent the possibility of tampering, the data repository can be replaced with a distributed file system. As an additional benefit, this makes it easier to recover the data. This, however, would lead to a complete change of the core product/system, something that is not desired.

\subsubsection{Security Issues}
Analyzing the AUDITEM's security is complex as it contains many components, and the implementation can be different. We can look at the security from the core functions of the verification task: immutable storage and data encryption. 
The immutable storage of the records and certificates is dependent on the choice of the underlying blockchain. In our implementation, the RAFT consensus protocol used in Hyperledger Fabric can tolerate $f$ number of crash nodes in a network with a total number of $n = 2f + 1$ nodes \cite{hyperledger}. This means that an organization can control the blockchain if it holds more than half of the nodes, which, of course, should be prevented at all times. More companies joining the consortium network reduces this probability. On the other hand, Hyperledger Fabric can hand out user certificates to prevent adversaries from misleading the network. However, the network can be compromised by an internal adversary. For example, an internal attacker sitting between DIVT and the database can send fake data to DIVT, causing the model to use the fake database to verify integrity. An accountability mechanism can partially prevent this issue.

For the comparison of the verification records, the SHA-256 hash function is used and can be considered secure \cite{gdprcompliantpersonaldata}. Besides, the identification attributes are stored publicly in our model; this is an intentional design since we assume those attributes do not contain sensitive information.
If for some reason, the identification attributes contain information that cannot be shared with the outside world, the decision could be made to use cryptographic hashes. In this case, users should be careful of dictionary attacks as identification attributes are often short identification tags \cite{pinkas2002securing}. 
To prevent the unauthorized share of data, the integrity verification records are protected using the AES encryption algorithm. The encryption ensures that information is only visible to designated partners who obtained the private keys. Of course, anyone can retrieve the encrypted integrity records, but the data is safe as long as the secret keys remain secret and the AES algorithm stays safe. This is based on the fact that in a properly implemented AES, there is currently no feasible attack that would allow someone without a key to access encrypted data \cite{biryukov2012cryptanalysis}. However, various implementations of AES have been reported to be potentially vulnerable to side-channel attacks and social engineering attacks \cite{osvik2006cache}. The use of the differential privacy technique can be effective in migrating side-channel attacks \cite{xu2018using}, but this inevitably reduces the data utility due to the introduction of data noise. Therefore, in this respect, the comparison and implementation of other alternative encryption algorithms are left for our future work.

\section{Conclusion}
\label{conclusion}
This paper proposes a novel decentralized data integrity verification model called AUDITEM. It fills the gaps in the existing literature by processing large-scale complex data stored in traditional data repositories without changing the system's core components. AUDITEM aims to replace the current expensive and untrustworthy third-party auditing services. In addition to decentralization, it is designed to achieve three objectives: scalability, private data protection, and GDPR compliance. 
A sub-module called DIVT is also developed to support automated operations and customizable verification operations. Experimental and analytical results show that AUDITEM is feasible and effective in meeting various business requirements for data integrity verification.

The future work can be directed in two main directions: blockchain platforms and cryptographic algorithms. In terms of the blockchain, we will continue to investigate other scalable and secure blockchain platforms for possible integration with AUDIEEM. Regarding cryptographic algorithms, the comparison and implementation of other alternative cryptographic algorithms will be studied in the future.

\ifCLASSOPTIONcaptionsoff
  \newpage
\fi

\section*{Acknowledgment}
This research is funded by the European Union's Horizon 2020 research and innovation program under grant agreements 825134 (ARTICONF project), 862409 (BlueCloud project) and 824068 (ENVRI-FAIR project). The research is also supported by the Chinese Scholarship Council, and EU LifeWatch ERIC.

\bibliographystyle{IEEEtran}
\bibliography{IEEEabrv,ref}

\begin{thebibliography}{10}
\providecommand{\url}[1]{#1}
\csname url@samestyle\endcsname
\providecommand{\newblock}{\relax}
\providecommand{\bibinfo}[2]{#2}
\providecommand{\BIBentrySTDinterwordspacing}{\spaceskip=0pt\relax}
\providecommand{\BIBentryALTinterwordstretchfactor}{4}
\providecommand{\BIBentryALTinterwordspacing}{\spaceskip=\fontdimen2\font plus
\BIBentryALTinterwordstretchfactor\fontdimen3\font minus
  \fontdimen4\font\relax}
\providecommand{\BIBforeignlanguage}[2]{{%
\expandafter\ifx\csname l@#1\endcsname\relax
\typeout{** WARNING: IEEEtran.bst: No hyphenation pattern has been}%
\typeout{** loaded for the language `#1'. Using the pattern for}%
\typeout{** the default language instead.}%
\else
\language=\csname l@#1\endcsname
\fi
#2}}
\providecommand{\BIBdecl}{\relax}
\BIBdecl

\bibitem{mork2015contemporary}
R.~Mork, P.~Martin, and Z.~Zhao, ``Contemporary challenges for data-intensive
  scientific workflow management systems,'' in \emph{Workflows in Support of
  Large-Scale Science Workshop (WORKS)}.\hskip 1em plus 0.5em minus 0.4em\relax
  ACM, 2015, pp. 1--11.

\bibitem{datarepository}
\BIBentryALTinterwordspacing
{Confluent, Inc.} (2021) Databases, data lakes, and data warehouses explained.
  Accessed 29 December 2021. [Online]. Available:
  \url{https://www.confluent.io/learn/databases-data-lakes-and-data-warehouses-compared/}
\BIBentrySTDinterwordspacing

\bibitem{boritz2005practitioners}
J.~E. Boritz, ``{IS} practitioners' views on core concepts of information
  integrity,'' \emph{Int. J. Account. Inf. Syst.}, vol.~6, no.~4, pp. 260--279,
  2005.

\bibitem{Spectra}
\BIBentryALTinterwordspacing
{California Spectra Instrumentation, Inc.} (2021) Data integrity verification.
  Accessed 25 December 2021. [Online]. Available:
  \url{https://spectralogic.com/features/data-integrity-verification/}
\BIBentrySTDinterwordspacing

\bibitem{ninglu}
N.~Lu, Y.~Zhang, W.~Shi, S.~Kumari, and K.-K.~R. Choo, ``A secure and scalable
  data integrity auditing scheme based on hyperledger fabric,'' \emph{Comput.
  Secur. J.}, vol.~92, p. 101741, 2020.

\bibitem{auditanalytics}
\BIBentryALTinterwordspacing
{Audit Analytics}. (2022) Trends in the european audit market. Accessed 18
  January 2022. [Online]. Available:
  \url{https://www.auditanalytics.com/doc/Audit-Analytics-Trends-in-the-European-Audit-Market.pdf}
\BIBentrySTDinterwordspacing

\bibitem{vademecum}
M.~Belotti, N.~Božić, G.~Pujolle, and S.~Secci, ``A vademecum on blockchain
  technologies: When, which, and how,'' \emph{IEEE Commun. Surv. Tutor.},
  vol.~21, no.~4, pp. 3796--3838, 2019.

\bibitem{kalis}
R.~Kalis and A.~Belloum, ``Validating data integrity with blockchain,'' in
  \emph{IEEE International Conference on Cloud Computing Technology and Science
  (CloudCom)}.\hskip 1em plus 0.5em minus 0.4em\relax IEEE, 2018, pp. 272--277.

\bibitem{brandon2016blockchain}
D.~Brandon, ``The blockchain: The future of business information systems,''
  \emph{Int. J. Acad. Bus. World}, vol.~10, no.~2, pp. 33--40, 2016.

\bibitem{zhou2018trustworthy}
H.~Zhou, C.~de~Laat, and Z.~Zhao, ``Trustworthy cloud service level agreement
  enforcement with blockchain based smart contract,'' in \emph{IEEE
  International Conference on Cloud Computing Technology and Science
  (CloudCom)}.\hskip 1em plus 0.5em minus 0.4em\relax IEEE, 2018, pp. 255--260.

\bibitem{9311629}
L.~Liu, J.~Feng, Q.~Pei, C.~Chen, Y.~Ming, B.~Shang, and M.~Dong,
  ``Blockchain-enabled secure data sharing scheme in mobile-edge computing: An
  asynchronous advantage actor–critic learning approach,'' \emph{IEEE
  Internet Things J.}, vol.~8, no.~4, pp. 2342--2353, 2021.

\bibitem{9548672}
H.~Yin, Z.~Zhang, J.~He, L.~Ma, L.~Zhu, M.~Li, and B.~Khoussainov, ``Proof of
  continuous work for reliable data storage over permissionless blockchain,''
  \emph{IEEE Internet Things J.}, pp. 1--1, 2021.

\bibitem{9698094}
T.~Li, H.~Wang, D.~He, and J.~Yu, ``Blockchain-based privacy-preserving and
  rewarding private data sharing for iot,'' \emph{IEEE Internet Things J.}, pp.
  1--1, 2022.

\bibitem{8936349}
M.~Zhaofeng, W.~Lingyun, W.~Xiaochang, W.~Zhen, and Z.~Weizhe,
  ``Blockchain-enabled decentralized trust management and secure usage control
  of iot big data,'' \emph{IEEE Internet Things J.}, vol.~7, no.~5, pp.
  4000--4015, 2020.

\bibitem{metje2020improved}
N.~Metje, A.~Hojjati, A.~Beck, and C.~D. Rogers, ``Improved underground
  utilities asset management--assessing the impact of the {UK} utility survey
  standard {(PAS128)},'' \emph{Proc. Inst. Civil Eng.-Munic. Eng.}, vol. 173,
  pp. 218--236, 2020.

\bibitem{chen2020dynamic}
X.~Chen, T.~Shang, F.~Zhang, J.~Liu, and Z.~Guan, ``Dynamic data auditing
  scheme for big data storage,'' \emph{Front. Comput. Sci.}, vol.~14, no.~1,
  pp. 219--229, 2020.

\bibitem{gan2018efficient}
Q.~Gan, X.~Wang, and X.~Fang, ``Efficient and secure auditing scheme for
  outsourced big data with dynamicity in cloud,'' \emph{Sci. China Inf. Sci.},
  vol.~61, no.~12, pp. 1--15, 2018.

\bibitem{hao2018decentralized}
K.~Hao, J.~Xin, Z.~Wang, Z.~Jiang, and G.~Wang, ``Decentralized data integrity
  verification model in untrusted environment,'' in \emph{Asia-Pacific Web
  (APWeb) and Web-Age Information Management (WAIM) Joint International
  Conference on Web and Big Data}.\hskip 1em plus 0.5em minus 0.4em\relax
  Springer, 2018, pp. 410--424.

\bibitem{wang2019blockchain}
H.~Wang, Q.~Wang, and D.~He, ``Blockchain-based private provable data
  possession,'' \emph{IEEE Trans. Dependable Secur. Comput.}, vol.~18, no.~5,
  pp. 2379--2389, 2021.

\bibitem{li2018blockchain}
R.~Li, T.~Song, B.~Mei, H.~Li, X.~Cheng, and L.~Sun, ``Blockchain for
  large-scale internet of things data storage and protection,'' \emph{IEEE
  Trans. Serv. Comput.}, vol.~12, no.~5, pp. 762--771, 2018.

\bibitem{zyskind2015decentralizing}
G.~Zyskind, O.~Nathan \emph{et~al.}, ``Decentralizing privacy: Using blockchain
  to protect personal data,'' in \emph{IEEE Security and Privacy Workshops
  (SPW)}.\hskip 1em plus 0.5em minus 0.4em\relax IEEE, 2015, pp. 180--184.

\bibitem{sun2020blockchain}
J.~Sun, X.~Yao, S.~Wang, and Y.~Wu, ``Blockchain-based secure storage and
  access scheme for electronic medical records in {IPFS},'' \emph{IEEE Access},
  vol.~8, pp. 59\,389--59\,401, 2020.

\bibitem{gangadevi2021survey}
K.~Gangadevi and R.~R. Devi, ``A survey on data integrity verification schemes
  using blockchain technology in cloud computing environment,'' in \emph{IOP
  Conference Series: Materials Science and Engineering}.\hskip 1em plus 0.5em
  minus 0.4em\relax IOP Publishing, 2021, p. 012011.

\bibitem{gdprcompliantpersonaldata}
N.~B. Truong, K.~Sun, G.~M. Lee, and Y.~Guo, ``{GDPR}-compliant personal data
  management: A blockchain-based solution,'' \emph{IEEE Trans. Inf. Forensic
  Secur.}, vol.~15, pp. 1746--1761, 2020.

\bibitem{wang2018blockchain}
S.~Wang, Y.~Zhang, and Y.~Zhang, ``A blockchain-based framework for data
  sharing with fine-grained access control in decentralized storage systems,''
  \emph{IEEE Access}, vol.~6, pp. 38\,437--38\,450, 2018.

\bibitem{varsha2014using}
B.~S. Varsha and P.~Suryateja, ``Using attribute-based encryption with advanced
  encryption standard for secure and scalable sharing of personal health
  records in cloud,'' \emph{Int. J. Comput. Sci. Inf. Technol.}, vol.~5, no.~5,
  pp. 6395--6399, 2014.

\bibitem{jeroen}
J.~Bergers, Z.~Shi, K.~Korsmit, and Z.~Zhao, ``{DWH-DIM}: A blockchain based
  decentralized integrity verification model for data warehouses,'' in
  \emph{2021 IEEE International Conference on Blockchain (Blockchain)}.\hskip
  1em plus 0.5em minus 0.4em\relax IEEE, 2021, pp. 221--228.

\bibitem{huang2020blockchain}
H.~Huang, J.~Lin, B.~Zheng, Z.~Zheng, and J.~Bian, ``When blockchain meets
  distributed file systems: An overview, challenges, and open issues,''
  \emph{IEEE Access}, vol.~8, pp. 50\,574--50\,586, 2020.

\bibitem{hyperledger}
\BIBentryALTinterwordspacing
{Hyperledger Foundation}. (2022) Hyperledger fabric documentation. Accessed 15
  January 2022. [Online]. Available:
  \url{https://hyperledger-fabric.readthedocs.io/en/release-2.2/}
\BIBentrySTDinterwordspacing

\bibitem{fabric_ecdsa}
\BIBentryALTinterwordspacing
{\vspace{0mm}Hyperledger Foundation}. (2021) Fabric {CA} user’s guide.
  Accessed 20 December 2021. [Online]. Available:
  \url{https://hyperledger-fabric-ca.readthedocs.io/en/release-1.4/users-guide.html}
\BIBentrySTDinterwordspacing

\bibitem{burr2003selecting}
W.~E. Burr, ``Selecting the advanced encryption standard,'' \emph{IEEE Secur.
  Priv.}, vol.~1, no.~2, pp. 43--52, 2003.

\bibitem{shi2019operating}
Z.~Shi, H.~Zhou, Y.~Hu, S.~Jayachander, C.~de~Laat, and Z.~Zhao, ``Operating
  permissioned blockchain in clouds: A performance study of hyperledger
  sawtooth,'' in \emph{2019 18th International Symposium on Parallel and
  Distributed Computing (ISPDC)}.\hskip 1em plus 0.5em minus 0.4em\relax IEEE,
  2019, pp. 50--57.

\bibitem{caliper}
\BIBentryALTinterwordspacing
{Hyperledger Foundation}. (2021) Hyperledger caliper. Accessed 21 December
  2021. [Online]. Available: \url{https://hyperledger.github.io/caliper/}
\BIBentrySTDinterwordspacing

\bibitem{metrics}
\BIBentryALTinterwordspacing
{\vspace{0mm}Hyperledger Foundation}. (2022) Hyperledger blockchain performance
  metrics white paper. Accessed 17 January 2022. [Online]. Available:
  \url{https://hyperledger-fabric.readthedocs.io/en/release-2.2/chaincode_lifecycle.html}
\BIBentrySTDinterwordspacing

\bibitem{li2020survey}
X.~Li, P.~Jiang, T.~Chen, X.~Luo, and Q.~Wen, ``A survey on the security of
  blockchain systems,'' \emph{Futur. Gener. Comp. Syst.}, vol. 107, pp.
  841--853, 2020.

\bibitem{pinkas2002securing}
B.~Pinkas and T.~Sander, ``Securing passwords against dictionary attacks,'' in
  \emph{ACM Conference on Computer and Communications Security (CCS)}.\hskip
  1em plus 0.5em minus 0.4em\relax ACM, 2002, pp. 161--170.

\bibitem{biryukov2012cryptanalysis}
A.~Biryukov and J.~Gro{\ss}sch{\"a}dl, ``Cryptanalysis of the full {AES} using
  {GPU}-like special-purpose hardware,'' \emph{Fundam. Inform.}, vol. 114, no.
  3-4, pp. 221--237, 2012.

\bibitem{osvik2006cache}
D.~A. Osvik, A.~Shamir, and E.~Tromer, ``Cache attacks and countermeasures: the
  case of {AES},'' in \emph{Cryptographers’ Track at the RSA Conference
  (CT-RSA)}.\hskip 1em plus 0.5em minus 0.4em\relax Springer, 2006, pp. 1--20.

\bibitem{xu2018using}
M.~Xu, A.~Papadimitriou, A.~Feldman, and A.~Haeberlen, ``Using differential
  privacy to efficiently mitigate side channels in distributed analytics,'' in
  \emph{European Workshop on Systems Security (EuroSec)}.\hskip 1em plus 0.5em
  minus 0.4em\relax ACM, 2018, pp. 1--6.

\end{thebibliography}

\end{document}